\documentclass[aps,floatfix,onecolumn,a4paper,noshowpacs, nofootinbib,superscriptaddress,11pt]{revtex4}
\usepackage{graphicx,float}\usepackage{graphicx,float}
\usepackage[all]{xy}
\usepackage{amsmath,upgreek}
\usepackage{amssymb}
\usepackage{color}
\usepackage{diagbox}
\usepackage{epsfig,bm}		
\usepackage{graphicx,epstopdf,bigints,upgreek}
\usepackage{subfigure}
\usepackage{pdfpages,slashed}
\usepackage[colorlinks,hyperindex]{hyperref}
\definecolor{green1}{RGB}{0,128,0}
\hypersetup{backref=true,pagebackref=true}
\hypersetup{%
  colorlinks = true,
  linkcolor  = blue,
  citecolor = cyan,
}
\usepackage{bookmark,textgreek}
\usepackage{hyperref,color,xcolor}
\hypersetup{hidelinks,hyperindex=true,colorlinks=true,breaklinks=true,urlcolor= blue}
\hypersetup{%
  colorlinks = true,
  linkcolor  = blue
}\usepackage{amssymb}
\newsavebox{\foobox}

\usepackage{graphicx,float,tikz}
\usepackage[all]{xy}
\newcommand\ringring[1]{%
  {
   \mathop{\kern0pt #1}\limits^{
     \vbox to-1.85ex{
       \kern-2ex 
       \hbox to 0pt{\hss\normalfont\kern.1em \r{}\kern-.45em \r{}\hss}%
       \vss 
     }
   }
  }
}\newcommand\orcidroldao{{\href{https://orcid.org/0000-0003-3978-532X}{\orcidicon}}}
\newcommand{\orcidicon}{%
	\begin{tikzpicture}
	\draw[lime, fill=lime] (0,0)
		circle [radius=0.16]
		node[white] {{\fontfamily{qag}\selectfont \tiny ID}};
	\draw[white, fill=white] (-0.0625,0.095)
		circle [radius=0.007];
	\end{tikzpicture}	\hspace{-2mm}
}

\newcommand{\bpartial}{\mathop{\partial\kern -4pt\raisebox{.8pt}{$|$}}}

\newcommand{\cvm}{current version of the manuscript}
\newcommand{\pvm}{previous version of the manuscript}

\newcommand{\bes}{\begin{subequations}}
\newcommand{\ees}{\end{subequations}}
\def\beq{\begin{eqnarray}}
 \newcommand{\clt}{\textcolor{black}}
  
\def\eeq{\end{eqnarray}}
\def\be{\begin{equation}}
\def\ee{\end{equation}}

\begin{document}

\title{Information entropy of nuclear electromagnetic transitions  in AdS/QCD}
\author{Roldao da Rocha\orcidroldao\!\!}
\email{roldao.rocha@ufabc.edu.br}
\affiliation{Federal University of ABC, Center of Mathematics, Santo Andr\'e, 09580-210, Brazil}
\begin{abstract}
Electromagnetic transitions involving nucleons and their resonances are here presented in the fermionic sector of the AdS/QCD soft-wall model, using the differential configurational entropy (DCE) as a measure of information entropy. The DCE is employed to derive adjustable parameters that define the minimal and nonminimal couplings in the nuclear interaction with a gauge vector field, in the nucleonic $\gamma N \to N^*(1535)$ transition. The values obtained comply with phenomenological data with good accuracy.
\end{abstract}
\maketitle
\section{Introduction}

The differential configurational entropy (DCE) is a measure of information entropy inherent to a physical system, accounting for correlations among wave modes, in momentum space, which portrays the spatial-complexity of the physical system \cite{Gleiser:2018kbq,Gleiser:2011di}. The DCE evaluates the compression rate of information into wave modes. For any physical system at a given macrostate, the DCE precisely measures the degree of ignorance associated with microstates, in momentum space, by reckoning the number of bits of additional information needed to specify it. The physical system can be represented by a probability distribution in a compressed data form, also coding information concerning the physical system. Higher values of DCE mean loss of information, therefore establishing a suitable criterion for the most prevalent and dominant wave states occupied by the physical system, also indicating domains of configurational stability attained at the DCE critical points \cite{Gleiser:2012tu}.
The DCE also measures the level of disorder of the physical system and is equivalent to the Gibbs--Boltzmann entropy, when the information dimension vanishes \cite{Gleiser:2018jpd,Sowinski:2015cfa,Bernardini:2016hvx}. The DCE has been also applied to quantum systems, comprising an information entropic measure quantifying quantum probability densities. The DCE evaluates the number of bits needed to designate the organization of a physical system. For computing it, a spatially-localized scalar field is required. One typically uses the energy density operator, as the temporal component of the energy-momentum tensor. Other choices, such as scattering amplitudes and cross-sections as well, are also suitable when fundamental interactions in QCD are studied \cite{Karapetyan:2018yhm,Karapetyan:2019fst,Karapetyan:epjp,Karapetyan:plb,Karapetyan:2020epl,Ma:2018wtw}.

The DCE has been systematically shown to be a very important criterion to examine some of the most essential phenomenological features of AdS/QCD, which has sparked the way to a thorough relevant set up to investigate the dynamics regulating strongly-coupled QFTs.
It has also brought a new formal and analytical understanding of the dynamics of confinement in QCD. Originally, AdS/CFT states that pure gravity in an AdS bulk of codimension one is dual to a flat conformal QFT at the AdS asymptotic boundary \cite{Aharony:1999ti,Witten:1998qj}.
Hence physical observables can be computed in a strongly-coupled gauge field theory, simply using gravity in the bulk \cite{Brodsky:2010ur,Brodsky:2014yha}. When one studies QCD as the gauge theory in the boundary, it is possible to emulate confinement when a soft infrared cutoff is introduced along the bulk fifth dimension, constructing the so-called AdS/QCD soft-wall models \cite{Vega:2008te,Karch:2006pv}. The bulk additional dimension corresponds to the energy scale $\Lambda_{\scalebox{.53}{\textsc{QCD}}}$. The AdS/QCD soft-wall model states that Regge trajectories rise from confinement in QCD when one generally uses a dilaton in the action that regulates the AdS bulk fields.
Besides the introduction of a background dilaton,
a warp factor in the AdS metric and the effective
potential of the action can be alternatively employed to generate soft-wall models \cite{Gutsche:2012bp}. Ref. \cite{Gutsche:2011vb} demonstrated that these procedures are equivalent in generating AdS/QCD soft-wall models, which can shed some new light on the mechanism of hadronic interactions, and the structure of nucleons, as well as the fundamental features of the electromagnetic transitions between nucleons and their resonances.
Recently, several relevant new results involving the mass spectra of glueballs, baryons, mesons, and their resonances, also at finite temperature, have been explained by the DCE, corroborating experimental data and lattice QCD \cite{daRocha:2021ntm,Ferreira:2020iry,Bazeia:2021stz,Braga:2021fey,Braga:2021zyi,Braga:2020hhs,Bernardini:2018uuy,Braga:2018fyc,Braga:2017fsb,Colangelo:2018mrt}. The mass spectra of higher-spin resonances in several meson and baryon families were also predicted and addressed by DCE techniques, complying with data in the Particle Data Group (PDG) \cite{pdg}.
In addition, nuclear reactions involving high-energy hadronic states were studied in the context of the DCE \cite{Karapetyan:2017edu,Karapetyan:2016fai}, also in the Color Glass Condensate of saturated gluon matter in QCD 
\cite{Karapetyan:2021crv,Karapetyan:2020yhs,Rougemont:2017tlu}. 
Other important aspects of the DCE were investigated in Refs. \cite{Correa:2015vka,Bazeia:2018uyg,Stephens:2019tav,Cruz:2018qby,Lee:2019tod}, also approaching important aspects of turbulence \cite{Barreto:2022ohl}.

Nucleonic resonances were investigated in the AdS/QCD soft-wall model, comprising
a power-scaling effective account of helicity amplitudes at large $Q^2$, with reasonable compliance with existing data at low and intermediate $Q^2$.
Nucleon resonances can be formulated in the AdS/QCD soft-wall in the context of the nucleon-Roper transition \cite{Gutsche:2017lyu}. To describe electromagnetic form factors of the nucleon and the Roper resonance, an extended version of the effective action of AdS/QCD has been implemented, when non-minimal terms are regarded in the action that governs the resonances \cite{Gutsche:2019yoo}. The electroexcitation of the $N^*(1535)$ resonance can be studied using the unitary isobar model MAID \cite{Drechsel:2007if}.
Such an approach allows for determining the longitudinal amplitude in the low $Q^2$ regime, whose virtual photons are sensitive to the outer nucleonic structure, being responsible for describing resonances in the paradigm of quark-core constituents. The $\gamma N \to N^*(1535)$ comprises a peculiar transition, since models that describe helicity transition amplitudes in the full range
of $Q^2$ lack, still 
\cite{Ramalho:2020nwk}.

This work is devoted to the investigation of the electromagnetic transitions between the nucleon and its resonances within the soft-wall AdS/QCD, particularly, to the study of the $N^*(1535)$ resonance. To accomplish it, the DCE will be employed, playing a fundamental role in deriving adjustable parameters that define the minimal and nonminimal couplings in the nuclear interaction with a gauge vector field, in the $\gamma N \to N^*(1535)$ transition. The derived parameters will be shown to corroborate experimental data with great accuracy, corresponding to the global minima of the DCE in three possible analyses, presenting the dominant states of higher configurational stability of the nucleonic system. This paper is organized as the following:
Sect. \ref{II} presents nucleon resonances in the AdS/QCD soft-wall model, also describing the interactions between the fermionic fields representing the nucleons and the gauge vector field, in the $\gamma N \to N^*(1535)$ transition.
In Sect. \ref{III}, after presenting the main features of the DCE computational protocol, the DCE underlying the $\gamma N \to N^*(1535)$ transition is computed for three possible cases, each time as a function of two among five adjustable parameters, in the AdS/QCD setup, that define the minimal and nonminimal couplings in the nuclear interaction. The derived parameters are shown to corroborate experimental data with great accuracy, corresponding to the global minima of the DCE, presenting the dominant states of higher configurational stability of the nucleonic system. In Sect. \ref{IV} the concluding remarks are presented, together with a more detailed discussion about the important results obtained. Some perspectives are also addressed.

\section{Nucleon resonances in the AdS/QCD soft-wall}
\label{II}
The soft-wall AdS/QCD can be employed to describe $\gamma N \to N^*(1535)$ transitions. Ref. \cite{Gutsche:2019yoo} introduced the electromagnetic minimal coupling of the nucleon and its $N^*(1535)$ resonance, based on the gauge-invariant coupling of two fermionic AdS fields with suitable twist-dimension. A satisfactory description of data regarding helicity amplitudes can be obtained, in the small-$Q^2$ regime \cite{Gutsche:2019yoo}. One of the key principles determining AdS/CFT is the isomorphism between the conformal group to the
 isometry group of the AdS bulk. In the soft-wall AdS/QCD formalism, the basic concept is the conformal Poincar\'e metric.
Denoting by $r$ the holographic coordinate along the AdS bulk, one possible coordinate system equipping AdS${}_5$ is
given, in the near-horizon limit, by
\beq ds^2 = \frac{L^2}{r^2} dr^2 + \frac{r^2}{L^2}(-dt^2 + dx_i dx_i),\qquad\qquad i = 1,2,3,\eeq which covers a Poincaré patch of a global AdS bulk, for $L$ being the radius of the AdS bulk, with boundary located at $r\to\infty$.
Conformal Poincaré coordinates can be then defined by $z = L^2/r$, yielding
\begin{equation}
g_{AB} \, x^A x^B = \epsilon^a_A \, \epsilon^b_B \,
\eta_{ab} \, x^A x^B = \frac{L^2}{z^2} \left(-dx_\mu dx^\mu + dz^2\right),\label{poin}
\end{equation}
with $g = |{\det}(g_{AB})| = z^{-10}$ being the magnitude
of the determinant of $g_{AB}$, and $\epsilon^a_A = \delta^a_A\,L/z$ is the vielbein. The boundary of the AdS bulk is located at $z = 0$. The $z\to\infty$ limit defines the Poincaré horizon, at which the warp factor $L^2/z^2$ approaches zero. One can immediately read off the scale invariance of the metric. The 5-dimensional coordinate $z$ encodes the energy scale of AdS/QCD. The metric (\ref{poin}) yields physical processes in AdS bulk, having alike (proper) energies, at distinct radial positions, to be related to energies that scale as $1/z$. It means that a physical process, in the boundary gauge theory, having a characteristic energy $\Lambda_{\scalebox{.53}{\textsc{QCD}}}$ is associated with an AdS bulk physical process located at $z \sim 1/\Lambda_{\scalebox{.53}{\textsc{QCD}}}$. Hence the ultraviolet regime $\Lambda_{\scalebox{.53}{\textsc{QCD}}}\to \infty$ regards $z\to0$, defining the near-boundary sector, whereas the infrared regime $\Lambda_{\scalebox{.53}{\textsc{QCD}}}\to0$ corresponds to $z\to\infty$, defining the near-horizon sector \cite{Peet:1998wn}.
Although QCD is not rigorously conformal, the QCD coupling slowly
varies at the small-momentum transfer regime. Hence, in the realm where the QCD coupling is almost constant, the masses of the involved quark masses can be disregarded, yielding QCD as a conformal gauge theory. Therefore confinement can be emulated when appropriate boundary conditions are imposed in the bulk cutoff
$z = z_0 \approxeq 1/\Lambda_{\scalebox{.53}{\textsc{QCD}}}$. The AdS metric can be altered to represent a
potential determining confinement \cite{Brodsky:2007hb}.

If one considers the nuclear resonances\footnote{The $p,n$ indexes refer to the proton and the neutron, respectively.} ($N^* = (N^*_{\scalebox{.6}{\textsc{$p$}}},N^*_{\scalebox{.6}{\textsc{$n$}}})$) within the AdS/QCD soft-wall approach, the action can be determined by the spin-$1/2$ fermionic field and a gauge vector.
The dilaton background field $\upphi(z) = {\mathsf{k}}^2 z^2$, where ${\mathsf{k}} = 383$ MeV is a universal scale parameter \cite{Gutsche:2012wb}, confines the vector fields in AdS space, also reproducing the Roper
mass $M = 1.440$ GeV, as well as the proton mass $M_{\scalebox{.5}{\textsc{Proton}}} = 938.272$ MeV \cite{Gutsche:2012wb}. Refs. \cite{Henningson:1998cd,Contino:2004vy,Hong:2006ta}
scrutinized fermionic sectors of AdS/CFT and derived the 2-point function for a fermionic spinor field
operator of scaling dimension
$\Delta = 2+|m|$, where $m$ is the 5-dimensional mass of the fermionic fields describing nucleons, which holds for the $s$-wave, $\ell=0$, approach. When states with $\ell\neq0$ are analyzed, a twist-dimension $\uptau$ must modify the conformal dimension as $\Delta\mapsto \Delta+\uptau$ \cite{Vega:2008te,BoschiFilho:2012xr,Branz:2010ub}.
The twist-dimension endowing the AdS fermionic field can be identified with the scaling dimension of the nucleon operator
$\uptau = \mathsf{N} + \mathsf{L}$, where $\mathsf{N}$ is the number of partons in the nucleon and $\mathsf{L}$ is the maximal value of the quark magnetic angular momentum, in the light-front wave function
\cite{Brodsky:2006uqa}.

The mechanism regulating the $N^*(1535)$ resonance takes into account the inclusion of the minimal coupling of the nucleon ($\ell=0$) and the $N^*(1535)$ resonance ($\ell=1$). Twist-dimensions for the nucleon and the $N^*(1535)$ resonance are $\uptau = 3$ and $\uptau = 4$, respectively. The leading twists of the nucleon and $N^*(1535)$ are linked with the photon, described by the gauge vector field, through the nonminimal coupling, since minimal couplings are forbidden by current conservation \cite{Gutsche:2019yoo}. The minimal coupling between the nucleon and the $N^*(1535)$ resonance is realized for equal twists, $\uptau_N = \uptau_{N^*(1535)} = 4$, for the leading electromagnetic coupling between nucleon and $N^*(1535)$, and supports to determine both the longitudinal and transverse helicity amplitudes, at low $Q^2$ \cite{Gutsche:2019yoo}. To couple the nucleon $N$ and the $N^*(1535)$ resonance to the vector field, one can represent the action
\beq
\label{actionS}
S &=& S_{\scalebox{.6}{\textsc{free}}} + S_{\scalebox{.6}{\textsc{int}}}\, \eeq
as the sum of a free part, regulating AdS fields dynamics under confinement,
\beq
S_{\scalebox{.6}{\textsc{free}}} &= \bigintss \sqrt{g} \, e^{-\upphi(z)} \,
\left( \mathfrak{L}_{\scalebox{.6}{\textsc{$N$}}}(x^\mu,z) + \mathfrak{L}_{\scalebox{.6}{\textsc{$N^*$}}}(x^\mu,z)
+ \mathfrak{L}_{\scalebox{.6}{\textsc{$V$}}}(x^\mu,z)\right)\,d^4x dz,
\eeq
and an action that governs interactions among the fermionic fields and the gauge vector field, given by
\beq
\label{actionS0}
S_{\scalebox{.6}{\textsc{int}}} &=& \int \sqrt{g} \, e^{-\upphi(z)} \,
\mathfrak{L}_{\scalebox{.6}{\textsc{int}}}(x^\mu,z)\,d^4x dz.
\eeq
Besides, denoting by ${\scalebox{.9}{\textsc{$\uppsi$}}}_{\uptau_\pm}(x^\mu,z)$ the bulk fermionic fields, which are eigenspinors of the chirality operator, with eigenvalue $+$ [$-$] corresponding to the right- and left-chirality
operators in the 4-dimensional boundary QCD, the $\mathfrak{L}_{\scalebox{.6}{\textsc{$N$}}}$, $\mathfrak{L}_{\scalebox{.6}{\textsc{$N^*$}}}$, $\mathfrak{L}_{\scalebox{.6}{\textsc{$V$}}}$ respectively denote the free interaction Lagrangians dictating the behavior of the nucleon, its associated resonance, and the gauge vector field, respectively reading \cite{Gutsche:2019yoo}
\beq
\label{actionS1}
\mathfrak{L}_{\scalebox{.6}{\textsc{$N$}}}(x^\mu,z) &=& \sum\limits_{\uptau; \upalpha=\pm} \, c_\uptau^{\scalebox{.6}{\textsc{$N$}}} \,
\bar{\scalebox{.9}{\textsc{$\uppsi$}}}^{\scalebox{.6}{\textsc{$N$}}}_{\upalpha,\uptau}(x^\mu,z) \, {}\slashed\partial_\upalpha(z) \, {\scalebox{.9}{\textsc{$\uppsi$}}}^{\scalebox{.6}{\textsc{$N$}}}_{\upalpha,\uptau}(x^\mu,z)
\,,\\
\label{actionS2}
\mathfrak{L}_{\scalebox{.6}{\textsc{$N^*$}}}(x^\mu,z) &=& \sum\limits_{\uptau; \upalpha=\pm} \, c_{\uptau+1}^{\scalebox{.6}{\textsc{$N^*$}}} \,
\bar{\scalebox{.9}{\textsc{$\uppsi$}}}^{\scalebox{.6}{\textsc{$N^*$}}}_{\upalpha,\uptau+1}(x^\mu,z) \, {}\slashed\partial_\upalpha(z) \, {\scalebox{.9}{\textsc{$\uppsi$}}}^{\scalebox{.6}{\textsc{$N^*$}}}_{\upalpha,\uptau+1}(x^\mu,z)
\,,\\
\label{actionS3}
\mathfrak{L}_{\scalebox{.6}{$V$}}(x^\mu,z) &=& - \frac{1}{4} V_{AB}(x^\mu,z)V^{AB}(x^\mu,z).
\eeq
The gauge vector field $V_A(x^\mu,z)$ determines the action (\ref{actionS3}), as its field strength is expressed as
 \beq
V_{AB} = \partial_A V_B - \partial_B V_A,\eeq 
and the Dirac operator in Eqs. (\ref{actionS1}, \ref{actionS2}) reads \cite{Gutsche:2017lyu,Gutsche:2019yoo}
\beq
\label{short1}
{}\slashed\partial_\pm(z) &=& \frac{i}{2} \Gamma^A
\! \stackrel{\leftrightarrow}{\partial}_{A} - \frac{i}{8}
\Gamma^A \omega_A^{ab} [\Gamma_a, \Gamma_b]
\, \mp \upphi(z) \mp L\mp \frac32\,,
\label{short}
\eeq
with spin connection  
\beq
\omega_A^{ab} = \frac1z\delta^{[a}_A \delta^{b]}_z,\eeq 
The Lagrangian densities describing the nucleon and the nucleonic resonance, respectively given by Eqs. (\ref{actionS1}) and (\ref{actionS2}) are the usual Dirac Lagrangian component, whereas the Lagrangian density (\ref{actionS3}) regards the conventional pure gauge field strength Lagrangian. The Lagrangian density responsible for the nuclear interaction with the vector field reads
\beq\label{actionS4}
\mathfrak{L}_{\scalebox{.6}{\textsc{int}}}(x^\mu,z) &=&
\sum\limits_{\uptau; \upalpha=\pm} \left( c_{\uptau+1}^{\scalebox{.6}{\textsc{$N^*N$}}} \,
\bar{\scalebox{.9}{\textsc{$\uppsi$}}}_{\upalpha,\uptau+1}^{\scalebox{.6}{\textsc{$N^*$}}}(x^\mu,z) {}{\mathsf{F}}^{\scalebox{.6}{\textsc{$N^*N$}}}_{{\scalebox{.6}{\textsc{min}}}}(x^\mu,z) \,
{\scalebox{.9}{\textsc{$\uppsi$}}}_{\upalpha,\uptau+1}^{\scalebox{.6}{\textsc{$N$}}}(x^\mu,z) \right.\nonumber\\&&\left.\qquad\quad+ \, d_\uptau^{\scalebox{.6}{\textsc{$N^*N$}}} \,
\bar{\scalebox{.9}{\textsc{$\uppsi$}}}_{\upalpha,\uptau+1}^{\scalebox{.6}{\textsc{$N^*$}}}(x^\mu,z) \, {}{\mathsf{F}}^{\scalebox{.6}{\textsc{$N^*N$}}}_{\upalpha,{\scalebox{.6}{\textsc{non-min}}}}(x^\mu,z) \,
{\scalebox{.9}{\textsc{$\uppsi$}}}_{\upalpha,\uptau}^{\scalebox{.6}{\textsc{$N$}}}(x^\mu,z)\right),
\eeq
which also contains Hermitian the respective conjugate terms,
where $c_{\uptau}^{\scalebox{.6}{\textsc{$N$}}}$, $c_{\uptau+1}^{\scalebox{.6}{\textsc{$N^*$}}}$,
$c_{\uptau+1}^{\scalebox{.6}{\textsc{$N^*N$}}}$, and $d_\uptau^{\scalebox{.6}{\textsc{$N^*N$}}}$ are adjustable parameters inducing a mixture for the contribution of AdS fields with different twist-dimension $\uptau\geq3$. The sums account for the $\upalpha = \pm$ chiralities. 
The notation  
\beq
\label{short}
{\mathsf{F}}^{\scalebox{.6}{\textsc{$N^*N$}}}_{{\scalebox{.6}{\textsc{min}}}}(x^\mu,z) &=& \mathcal{Q} \, \Gamma^A V_A(x^\mu,z)\,,\\
\label{short2}
{}{\mathsf{F}}^{\scalebox{.6}{\textsc{$N^*N$}}}_{\pm,{\scalebox{.6}{\textsc{non-min}}}}(x^\mu,z) &=& \pm \,
\frac{i}{4} \eta_{\scalebox{.6}{\textsc{$V$}}}^{\scalebox{.6}{\textsc{$N^*N$}}} \sigma^{AB} V_{AB}(x^\mu,z) + \zeta_{\scalebox{.6}{\textsc{$V$}}}^{\scalebox{.6}{\textsc{$N^*N$}}} \, z \, \Gamma^A \, \partial^B V_{AB}(x^\mu,z)
\,
\eeq
regards the Lagrangian density (\ref{actionS4}), with $\Gamma^A = \epsilon^A_a \Gamma^a$ and
$\Gamma^a = (\upgamma^\mu, \upgamma^0\upgamma^1\upgamma^2\upgamma^3)$, for $\upgamma^\mu$ being the gamma Dirac matrices, and $\sigma^{AB}= \left[\Gamma^A, \Gamma^B\right]$.
In Eqs. (\ref{actionS4}, \ref{short}, \ref{short2}) the subscripts ${\scalebox{.6}{\textsc{min}}}$ and ${\scalebox{.6}{\textsc{non-min}}}$ respectively refer
 to minimal and non-minimal couplings, and $\mathcal{Q} = {\scalebox{.9}{\textsc{diag}}}(1,0)$ represents the charge matrix of the nucleonic resonance $N^*$ \cite{Gutsche:2011vb,Gutsche:2012bp}. 

It is worth emphasizing that the relevant developments in the literature
also take into account, besides $\uptau = 3$ corresponding to nucleon with three quarks-component, the twist-4 Fock states ($\uptau = 4)$, with three quarks and one gluon, and also twist-5 Fock states ($\uptau = 5)$, containing three quarks and a $q\bar{q}$ component \cite{Gutsche:2012wb,Gutsche:2019yoo}. These are the three leading twist contributions to the $N^*(1535)$ resonance mass, corresponding to the twist-dimension $\uptau_{\scalebox{.6}{\textsc{$N^*$}}}\in\{4,5,6\}$. Their expression can be derived when the equations of motion for the fermionic Kaluza--Klein modes are constructed, with a consistent asymptotic behavior of the nucleon electromagnetic form factors at large $Q^2$. 
The values of the couplings $\eta_{\scalebox{.6}{\textsc{$V$}}}$ and $\zeta_{\scalebox{.6}{\textsc{$V$}}}^{\scalebox{.6}{\textsc{$N^*N$}}}$ are fixed from the magnetic moments, slopes, and form factors of the nucleons and their respective resonances, being derived in Refs. \cite{Gutsche:2017lyu,Gutsche:2019yoo} as
\beq\eta_{\scalebox{.6}{\textsc{$V$}}}^{\scalebox{.6}{\textsc{$N^*N$}}} = 4.28\,, \qquad
 \zeta_{\scalebox{.6}{\textsc{$V$}}}^{\scalebox{.6}{\textsc{$N^*N$}}} = -0.47\,.\label{cou1}\eeq
 There are some restrictions on the minimal coupling between the nucleon and its resonance. Regarding gauge invariance, the nucleon and its resonance have the same twist-dimension. In the case of the nonminimal coupling, however, there are other options for the coupling modes, including the coupling between the nucleon with a twist-dimension $\uptau_{\scalebox{.6}{\textsc{$N$}}} \in\{3,4,5\}$ and the $N^*(1535)$ resonance with the twist-dimension $\uptau_{\scalebox{.6}{\textsc{$N^*$}}} = \uptau_{\scalebox{.6}{\textsc{$N$}}}+1$.
 
The AdS bulk fermionic fields for the nucleon and its resonance have been denoted, respectively, by ${\scalebox{.9}{\textsc{$\uppsi$}}}^{\scalebox{.6}{\textsc{$N$}}}_{\uptau_\pm}(x^\mu,z)$ and ${\scalebox{.9}{\textsc{$\uppsi$}}}^{\scalebox{.6}{\textsc{$N^*$}}}_{\uptau_\pm}(x^\mu,z)$.
These fermionic fields are dual to the left-handed and
right-handed doublets representing the nucleon and the $N^*(1535)$ resonance, where 
${\cal O}^{\scalebox{.6}{\textsc{$L$}}} = \left(B_{\scalebox{.6}{\textsc{$p$}}}^{\scalebox{.6}{\textsc{$L$}}}, B_{\scalebox{.6}{\textsc{$n$}}}^{\scalebox{.6}{\textsc{$L$}}}\right)^\intercal$ and ${\cal O}^{\scalebox{.6}{\textsc{$R$}}} = \left(B_{\scalebox{.6}{\textsc{$p$}}}^{\scalebox{.6}{\textsc{$R$}}}, B_{\scalebox{.6}{\textsc{$n$}}}^{\scalebox{.6}{\textsc{$R$}}}\right)^\intercal$
where $B_{\scalebox{.6}{\textsc{$p$}}}$ regards the proton and the protonic $N^*_{\scalebox{.6}{\textsc{$p$}}}$ resonance, whereas $B_{\scalebox{.6}{\textsc{$n$}}}$ denotes the neutron and its resonance $N^*_{\scalebox{.6}{\textsc{$n$}}}$. The fermionic fields can be determined as the products of the left- and right-handed boundary spinor fields, respectively for the nucleon and resonances, as
\begin{equation}
\label{psi_expansion_Nucleon}
{\scalebox{.9}{\textsc{$\uppsi$}}}^{\scalebox{.6}{\textsc{$L$}}}_{\scalebox{.6}{\textsc{$N$}}}(x^\mu)
 = \frac12\left(\mathbb{I}- \upgamma^5\right){\scalebox{.9}{\textsc{$\uppsi$}}}(x^\mu),
 \qquad\qquad {\scalebox{.9}{\textsc{$\uppsi$}}}^{\scalebox{.6}{\textsc{$R$}}}_{\scalebox{.6}{\textsc{$N$}}}(x^\mu)
 = \frac12\left(\mathbb{I}+ \upgamma^5\right){\scalebox{.9}{\textsc{$\uppsi$}}}(x^\mu),
\end{equation}
and
\begin{equation}
\label{psi_expansion_Nstar}
{\scalebox{.9}{\textsc{$\uppsi$}}}^{\scalebox{.6}{\textsc{$L$}}}_{\scalebox{.6}{\textsc{$N^*$}}}(x^\mu) = -\frac12 \left({\mathbb{I}+ \upgamma^5}\right){\scalebox{.9}{\textsc{$\uppsi$}}}(x^\mu),\qquad\qquad {\scalebox{.9}{\textsc{$\uppsi$}}}^{\scalebox{.6}{\textsc{$R$}}}_{\scalebox{.6}{\textsc{$N^*$}}}(x^\mu) = \frac12 \left({\mathbb{I}+ \upgamma^5}\right){\scalebox{.9}{\textsc{$\uppsi$}}}(x^\mu).
\end{equation}
The fermionic bulk profiles can be split into the pure boundary and pure bulk components, emulating a Kaluza--Klein expansion as
\beq
\label{psi_expansion2}
{\scalebox{.9}{\textsc{$\uppsi$}}}^{\scalebox{.6}{\textsc{$N$}}}_{\uptau_\pm}(x^\mu,z) &=& \frac{\sqrt{2}}{{2}} \,
\left[
\pm {\scalebox{.9}{\textsc{$\uppsi$}}}^{\scalebox{.6}{\textsc{$L$}}}_{\scalebox{.6}{\textsc{$N$}}}(x^\mu) \ F^{\scalebox{.6}{\textsc{$L,R$}}}_{\uptau}(z)
+ {\scalebox{.9}{\textsc{$\uppsi$}}}^{\scalebox{.6}{\textsc{$R$}}}_{\scalebox{.6}{\textsc{$N$}}}(x^\mu) \ F^{\scalebox{.6}{\textsc{$R,L$}}}_{\uptau}(z)\right]\,,\\
\label{psi_expansion2}
{\scalebox{.9}{\textsc{$\uppsi$}}}^{\scalebox{.6}{\textsc{$N^*$}}}_{\uptau_\pm}(x^\mu,z) &=& \frac{\sqrt{2}}{{2}} \,
\left[
\mp {\scalebox{.9}{\textsc{$\uppsi$}}}^{\scalebox{.6}{\textsc{$L$}}}_{\scalebox{.6}{\textsc{$N^*$}}}(x^\mu) \ F^{\scalebox{.6}{\textsc{$L,R$}}}_{\uptau}(z)
+ {\scalebox{.9}{\textsc{$\uppsi$}}}^{\scalebox{.6}{\textsc{$R$}}}_{\scalebox{.6}{\textsc{$N^*$}}}(x^\mu) \ F^{\scalebox{.6}{\textsc{$R,L$}}}_{\uptau}(z)\right]\,,
\eeq
where the normalizable profile functions $F^{\scalebox{.6}{\textsc{$R,L$}}}_{\uptau}(z)$ are the holographic counterpart of nucleonic wave functions with specific twist-dimension and are given by
\begin{equation}
\label{psi_expansion1}
F^{\scalebox{.6}{\textsc{$L,R$}}}_{\uptau}(z) = z^2e^{-{\mathsf{k}}^2 z^2/2} \, f^{\scalebox{.6}{\textsc{$L,R$}}}_{\uptau}(z),
\end{equation}
and
\beq
\label{fL_fR}
f^{\scalebox{.6}{\textsc{$L$}}}_{\uptau}(z) &=& \sqrt{{2}}{\left(\Upgamma(\uptau)\right)^{-1/2}} \, {\mathsf{k}}^{\uptau} \,
z^{\uptau - 1/2}\,,\\
\label{fL_fR}
f^{\scalebox{.6}{\textsc{$R$}}}_{\uptau}(z) &=& \sqrt{{2}}{\left(\Upgamma(\uptau-1)\right)^{-1/2}} \, {\mathsf{k}}^{\uptau-1} \,
z^{\uptau - \frac32}\,.
\eeq
It is worth emphasizing that the ground state of the nucleon is characterized by $n=\ell=0$, whereas the $N^*(1535)$ resonance $n=0$ consists of an excited state $\ell=1$.
For the gauge vector field, the axial gauge $V_z = 0$ can be regarded. An inverse Fourier transformation with respect to the boundary coordinates is implemented as
\begin{equation}
\label{V_Fourier}
V_\mu(x^\mu,z) = \frac{1}{(2\pi)^2}\int_{\mathbb{R}^{1,3}} e^{iq\cdot x}\,V_\mu(q) V(q,z)\,d^4q.
\end{equation}
Once the Fourier transform \eqref{V_Fourier} of the gauge vector field is calculated, the equation of motion that regulates the vector bulk-to-boundary propagator concerns the $q^2$-dependent electromagnetic current as
\begin{equation}
\label{eq}
\partial_z \left( \frac{e^{-\upphi(z)}}{z} \,
\partial_z V(q,z)\right) + \frac{q^2}{z} e^{-\upphi(z)}
V(q,z) = 0,
\end{equation}
whose analytical solutions involve the
gamma $\Upgamma(n)$ and Tricomi $U(a,b,z)$ functions,
\begin{equation}
\label{VInt_q}
V(q,z) = \Upgamma\left(1 - \frac{q^2}{4{\mathsf{k}}^2}\right)
\, U\left(-\frac{q^2}{4{\mathsf{k}}^2},0,{\mathsf{k}}^2 z^2\right) \,.
\end{equation}
The bulk-to-boundary propagator $V(q,z)$ obeys the normalization condition $\lim_{q\to0}V(q,z) = 1$, complying to gauge invariance, also satisfying the infrared and ultraviolet boundary conditions
\beq
 \lim_{z\to\infty} V (q,z) = 0,\qquad\qquad \lim_{z\to0}V(q, z) = 1.\label{cond}
\eeq
The last condition in (\ref{cond}) regards the local coupling of the electromagnetic field to fermionic fields, whereas the first one reflects the vanishing vector field at $z\to\infty$.
The integral representation for the solution of Eq. (\ref{eq}) at the Euclidean mode ($Q^2 = - q^2 > 0$) is useful for deriving analytical expressions, reading \cite{Brodsky:2007hb,Grigoryan:2007my}
\begin{equation}
\label{VInt}
V(Q,z) = {\mathsf{k}}^2 z^2 \int_0^1 \frac{x^{a}}{(1-x)^2}
\exp\left(- {\mathsf{k}}^2 z^2 \frac{x}{1-x}\right)\,dx,
\end{equation}
denoting the light-cone momentum fraction by $x$ and 
$a = Q^2/(4 {\mathsf{k}}^2)$.
In the Lagrangian densities (\ref{actionS1}, \ref{actionS2}), governing the 4-dimensional boundary fermionic fields, the parameters $c_\uptau^{\scalebox{.6}{\textsc{$N$}}}$ and $c_{\uptau+1}^{\scalebox{.6}{\textsc{$N^*$}}}$ are normalized as $\sum_\uptau \, c_\uptau^{\scalebox{.6}{\textsc{$N$}}} = 1$
and $\sum_\uptau \, c_{\uptau+1}^{\scalebox{.6}{\textsc{$N^*$}}} = 1$, arising from the consistent normalization of the fermionic kinetic term in the Dirac equation and also from the electromagnetic U(1) gauge invariance \cite{Gutsche:2012wb,Gutsche:2012bp}.
As the normalization condition complies with gauge invariance, the nucleon mass, and the $N^*(1535)$ resonance mass can be written as a linear combination of the square root of the twist-dimension,
\begin{equation}
\label{lincom}
M_{\scalebox{.6}{\textsc{$N$}}} = 2 {\mathsf{k}} \sum\limits_\uptau\, c_\uptau^{\scalebox{.6}{\textsc{$N$}}}\, \sqrt{\uptau - 1}
\,, \qquad\qquad\quad
M_{\scalebox{.6}{\textsc{$N^*$}}} = 2 {\mathsf{k}}
\sum\limits_\uptau \, c_{\uptau+1}^{\scalebox{.6}{\textsc{$N^*$}}}\, \sqrt{\uptau}\,.
\end{equation}
The coefficients $c_\uptau^{\scalebox{.6}{\textsc{$N$}}}$ and $c_{\uptau+1}^{\scalebox{.6}{\textsc{$N^*$}}}$ of the linear combinations \eqref{lincom} are the same as the ones entering the fermionic Lagrangian densities (\ref{actionS1}, \ref{actionS2}), respectively.
Regarding the mass of the $N^*(1535)$ resonance, the model is limited by the
three leading twist contributions, $(\uptau_{\scalebox{.6}{\textsc{$N^*$}}} \in \{4,5,6\})$.
For the value of the $N^*(1535)$ mass equal to 1.510 GeV, Refs. \cite{Gutsche:2012wb,Gutsche:2019yoo} derived the values \beq
c_4^{\scalebox{.6}{\textsc{$N^*$}}} = 0.82, \quad\quad\qquad c_5^{\scalebox{.6}{\textsc{$N^*$}}} = -0.63,\eeq
dictating the free Lagrangian density (\ref{actionS2}) for the nucleonic resonance.
As the model includes the $\sum_\uptau \, c_{\uptau+1}^{\scalebox{.6}{\textsc{$N^*$}}} = 1$ normalization, the parameters $c_4^{\scalebox{.6}{\textsc{$N^*$}}} $ and $c_5^{\scalebox{.6}{\textsc{$N^*$}}} $ are linearly independent,
yielding \beq
c_6^{\scalebox{.6}{\textsc{$N^*$}}} = 1 - c_4^{\scalebox{.6}{\textsc{$N^*$}}} - c_5^{\scalebox{.6}{\textsc{$N^*$}}} = 0.81.\eeq
The following parameters, entering the Lagrangian density \eqref{actionS4} governing the nuclear fermionic interaction with the gauge vector field, encoded in the Lagrangian density \eqref{actionS4}, were derived from a fit involving the transversal and longitudinal helicity amplitudes of the $\gamma N \to N^*(1535)$ transition \cite{Gutsche:2019yoo},
\beq \!\!\!\!\!\!c_4^{\scalebox{.6}{\textsc{$N^*N$}}} = 25.52\,, \qquad\quad\quad
 c_5^{\scalebox{.6}{\textsc{$N^*N$}}} = -26.90\,\label{par1}\eeq
 and
 \beq
d_3^{\scalebox{.6}{\textsc{$N^*N$}}} = -1.89\,, \quad\qquad\quad
 d_4^{\scalebox{.6}{\textsc{$N^*N$}}} = 5.64\,, \quad\qquad\quad
 d_5^{\scalebox{.6}{\textsc{$N^*N$}}} = -3.58\,.\label{par2}
\eeq
In the next section, two among these parameters will be assumed free, fixing the other three parameters, and the DCE will be employed to derive the values of the free parameters which correspond to the maximal configurational stability of the nuclear system. In other words, the best choice of the free parameters corresponds to the global minima of the DCE, for each case analyzed involving a mixture of two parameters among the ones in Eqs. (\ref{par1}, \ref{par2}).

\clt{To perform the computations using QFT methods together with the known form factors, in the effective theories describing low energy limit of QCD, one can  scrutinize the  form factors and helicity 
amplitudes of the $\gamma N \to N^*(1535)$ transition. 
The electromagnetic form factors of the $\gamma N \to N^*(1535)$ transition 
are encoded by  the gauge invariant object
\beq\label{matrix_elements}
\!\!\!\!\!\!\!\!\!M^\mu(p_1\lambda_1,p_2\lambda_2) &=& \bar u_{N^*}(p_1\lambda_1)
\biggl[ \gamma^\mu_\perp \, F_1^{N^*N}(-q^2) 
+ i \sigma^{\mu\nu} \frac{q_\nu}{M_{N^*} + M_N} \, F_2^{N^*N}(-q^2) 
\, \biggr] \, \gamma^5 \, u_{N}(p_2\lambda_2)\,,
\eeq
where $u_{N^*}(p_1,\lambda_1)$ and $u_{N}(p_2,\lambda_2)$ represent Dirac spinors describing the 
$N^*(1535)$ resonance and the nucleon,  $\gamma^\mu_\perp = \gamma^\mu 
- q^\mu \not\! q/q^2$\,, 
$q = p_1 - p_2$, whereas $\lambda_1$, $\lambda_2$, and $\lambda$ are 
the helicities of the final, initial baryon and the photon, respectively, 
satisfying the identity $\lambda_2 = \lambda_1 - \lambda$. 
The term  
$F_1^{N^*N}(Q^2)$, appearing in nonminimal terms of the actions~(\ref{actionS}, \ref{actionS0}), includes the derivative with respect to the $z$ coordinate, acting on the bulk-to-boundary 
propagator $\partial_z V(Q,z)$. Regarding minimal terms, one can write 
\beq \label{form1}
F_{1}^{N^*N}(Q^2) = \frac{a}{2} \, 
\sum\limits_{\tau} \, c_{\tau+1}^{N^*N} \,  \frac{\Gamma(a+1) \Gamma(\tau+1)}{\Gamma(a+\tau+2)}\,,
\eeq  
One can realize that the two parameters in Eq. (\ref{par1}) appear in Eq. (\ref{form1}). However, to assess the other parameters in the model, exclusively using QFT techniques, one must introduce the nucleon-Roper transition, as in Ref. \cite{Gutsche:2019yoo}. Besides, Ref. \cite{Konen:1989jp} addressed the electromagnetic $N \to N^*$ (1535) transition in the relativistic constituent quark model, using two parameters, but the range of $Q^2$ is pretty limited. Ref. \cite{Jido:2007sm} introduced the way how electromagnetic properties yield  useful insights about the
$N^*(1535)$ transition form factors, which are dynamically generated from the strong interaction regarding mesons and baryons. However, this QFT approach for the electromagnetic $N \to N^*(1535)$ transition does not contain explicitly non-minimal terms. Other approaches in the literature in Refs. \cite{Gutsche:2012wb,Gutsche:2017lyu,Gutsche:2012bp} consider also the Roper resonance. Ref. \cite{Tiator:2011pw} approaches phenomenological and experimental aspects of electromagnetic properties of nucleon resonance excitation and the number of parameters is lower, as long as non-minimal terms are not included. Refs. \cite{Aznauryan:2011qj,Aznauryan:2012ec} also discuss nucleon electromagnetic form factors and electroexcitation of low-lying nucleon
resonances in a light-front relativistic quark model, using the same number of parameters as the AdS/QCD approach. However, in the AdS/QCD setup, the inclusion of non-minimal terms is more natural, as discussed in Ref. \cite{Gutsche:2019yoo} and used in our work. 
}

\section{DCE underlying nucleon resonances in the AdS/QCD soft-wall}
\label{III}

The DCE protocol consists, first, to consider a probability distribution, which is usually taken as the localized energy density operator -- the temporal $\uptau_{00}({\mathsf{r}})$ component of the stress-energy-momentum tensor, for ${\mathsf{r}}=(x_1,\ldots,x_p)\in\mathbb{R}^p$. The 2-point correlation function,
\beq
\Uppi({\mathsf{r}})=\int_{\mathbb{R}^p} \, \uptau_{00}(\vec{r})\uptau_{00}(\vec{r}+{\mathsf{r}})\,d\vec{r}\label{corr}\eeq
 establishes the DCE as the information entropy of correlations among the wave modes composing the nuclear system, since the correlation function measures order and the way how microscopic variables vary with one another, on average \cite{Ma:2018wtw,Braga:2018fyc}. The 2-point correlation function \eqref{corr} also measures fluctuations of the energy density and the propensity of the nuclear system to homogenize itself.
 Now denoting by ${\mathsf{q}}$ the spatial part of the 4-momentum $q$, the protocol to derive the DCE has a first step of computing the Fourier transform of the energy density operator \cite{Gleiser:2018kbq},
\beq\label{fou}
\uptau_{00}({\mathsf{q}}) = \frac{1}{(2\pi)^{p/2}}\int_{\mathbb{R}^p}\,\uptau_{00}({\mathsf{r}})e^{-i{\mathsf{q}}\cdot {\mathsf{r}}}\,{\rm d}^p x,\eeq
representing the normalized spectral density per unit volume. 
It emulates the definition of collective coordinates in statistical mechanics \cite{Bernardini:2016hvx}, as $\uptau_{00}({\mathsf{r}})$ is the energy density operator. Now, wave modes in a $p$-volume ${\rm d}^p{q}$, centered at ${\mathsf{q}}$, can be related to a probability that encodes the spectral density associated with the wave
modes \cite{Gleiser:2018kbq},
\beq\label{sd}
\mathsf{P}\left({\mathsf{q}}\,\vert\, {\rm d}^p\mathsf{q}\right)\propto \left|\uptau_{00}({\mathsf{q}})\right|^{2}{\rm d}^p\mathsf{q}.
\eeq The Fourier transform (\ref{fou}) and the spectral density (\ref{sd}) motivate the definition of the modal fraction
\cite{Gleiser:2012tu},
\begin{eqnarray}
\tau_{00}({\mathsf{q}}) = \frac{\left|\uptau_{00}({\mathsf{q}})\right|^{2}}{ \bigintsss_{\mathbb{R}^p} \,\left|\uptau_{00}(\mathfrak{q})\right|^{2}\,{\rm d}^p\mathfrak{q}}.\label{modalf}
\end{eqnarray}
It mirrors the concept of structure factor in statistical mechanics, encoding the relative weight transported by each momentum wave mode ${\mathsf{q}}$.
The weight of each wave mode, representing information in the nuclear system, that is demanded to express the energy density, can be computed by the DCE \cite{Gleiser:2018kbq},
\begin{eqnarray}
{\rm DCE}_{\tau_{00}}= - \int_{\mathbb{R}^p}\,{\tau_{00}^{\scalebox{.57}{$\,\ominus$}}}({\mathfrak{q}})\ln {\tau_{00}^{\scalebox{.57}{$\,\ominus$}}}({\mathfrak{q}})\,{\rm d}^p\mathfrak{q},
\label{confige}
\end{eqnarray}
for \beq{\tau_{{00}}^{\scalebox{.57}{$\,\ominus$}}({\mathsf{q}})=\frac{\tau_{00}({\mathsf{q}})}{\tau_{{00}}^{\scalebox{.58}{max}}({\mathsf{q}})}},\eeq and ${\tau_{{00}}^{\scalebox{.58}{max}}({\mathsf{q}})}$ denoting the maximum value of the energy density in the momentum space $\mathbb{R}^p$. Besides, in the state ${\tau_{{00}}^{\scalebox{.58}{max}}({\mathsf{q}})}$ the spectral density is the highest one. DCE units are several, however, nat (natural unit of information) is the most used, quantifying the amount of information underlying a probability distribution, in which all outcomes are equally likely, in the interval $[0,e]$. The DCE, in the context presented in Sect. \ref{II}, is a global outspread measure of the nucleonic state density.

The DCE of the nucleonic system can be computed by fixing $p=1$ in (\ref{fou}) -- (\ref{confige}), since all the integrations involved are calculated along the holographic coordinate $z$, due to the Kaluza--Klein splitting. Regarding the Lagrangian densities (\ref{actionS1}) -- (\ref{actionS4})
and denoting by
\beq\label{full}
\mathfrak{L} =
\mathfrak{L}_{\scalebox{.6}{\textsc{$N$}}}+\mathfrak{L}_{\scalebox{.6}{\textsc{$N^*$}}} +\mathfrak{L}_{\scalebox{.6}{$V$}} +\mathfrak{L}_{\scalebox{.6}{\textsc{int}}},
\eeq one can substitute the Lagrangian density (\ref{full}) into the expression for the energy density operator, namely, the temporal component of the stress-energy-momentum tensor,
 \begin{equation}
\!\!\!\!\!\!\!\!\uptau_{00}\!=\! \frac{2}{\sqrt{ -g }}\!\! \left[\frac{\partial (\sqrt{-g}{\mathfrak{L}})}{\partial{g^{00}}} \!-\!\frac{\partial}{\partial{ x^\rho }} \frac{\partial (\sqrt{-g} {\mathfrak{L}})}{\partial\left(\frac{{\scalebox{.79}{$\,\partial$}} g^{00}}{{\scalebox{.79}{$\,\partial$}}x^\rho}\right)} \right]+\partial_0\bar\psi\frac{\partial {\mathfrak{L}}}{\partial\left(\partial^0\bar\psi\right)}+\frac{\partial {\mathfrak{L}}}{\partial\left(\partial^0\psi\right)}\partial_0\psi - g_{00}{\mathfrak{L}},
 \label{em1}
 \end{equation}
\clt{ where $\psi$ accounts for any fermionic field entering in the Lagrangians \eqref{actionS1}, \eqref{actionS2}, and \eqref{actionS4}.
Explicitly, Eq. (\ref{em1}) can be written as 
\beq
\label{actionS55}
\uptau_{00}\! &\!=\!&\!\! \sum\limits_{\uptau; \upalpha=\pm} \left[\, c_\uptau^{\scalebox{.6}{\textsc{$N$}}} \,
\bar{\scalebox{.9}{\textsc{$\uppsi$}}}^{\scalebox{.6}{\textsc{$N$}}}_{\upalpha,\uptau}(x^\mu,z) \, \gamma_0{}\overset{\leftrightarrow}{\partial_{\upalpha}}(z) \, {\scalebox{.9}{\textsc{$\uppsi$}}}^{\scalebox{.6}{\textsc{$N$}}}_{\upalpha,\uptau}(x^\mu,z)+ c_{\uptau+1}^{\scalebox{.6}{\textsc{$N^*$}}} \,
\bar{\scalebox{.9}{\textsc{$\uppsi$}}}^{\scalebox{.6}{\textsc{$N^*$}}}_{\upalpha,\uptau+1}(x^\mu,z) \gamma_0{}\overset{\leftrightarrow}{\partial_{\upalpha}}(z) \, {\scalebox{.9}{\textsc{$\uppsi$}}}^{\scalebox{.6}{\textsc{$N^*$}}}_{\upalpha,\uptau+1}(x^\mu,z)\right.\nonumber\\
&&\left.+g_{00}\left(c_{\uptau+1}^{\scalebox{.6}{\textsc{$N^*N$}}} \,
\bar{\scalebox{.9}{\textsc{$\uppsi$}}}_{\upalpha,\uptau+1}^{\scalebox{.6}{\textsc{$N^*$}}}(x^\mu,z) {}{\mathsf{F}}^{\scalebox{.6}{\textsc{$N^*N$}}}_{{\scalebox{.6}{\textsc{min}}}}
{\scalebox{.9}{\textsc{$\uppsi$}}}_{\upalpha,\uptau+1}^{\scalebox{.6}{\textsc{$N$}}}(x^\mu,z)+d_\uptau^{\scalebox{.6}{\textsc{$N^*N$}}} \,
\bar{\scalebox{.9}{\textsc{$\uppsi$}}}_{\upalpha,\uptau+1}^{\scalebox{.6}{\textsc{$N^*$}}}(x^\mu,z){\mathsf{F}}^{\scalebox{.6}{\textsc{$N^*N$}}}_{\upalpha,{\scalebox{.6}{\textsc{non-min}}}}
{\scalebox{.9}{\textsc{$\uppsi$}}}_{\upalpha,\uptau}^{\scalebox{.6}{\textsc{$N$}}}(x^\mu,z)\right)\right]\nonumber\\
&&- \frac{1}{4}g_{00} V_{AB}(x^\mu,z)V^{AB}(x^\mu,z)+V_{0A}(x^\mu,z)V_0^{A}(x^\mu,z).\eeq
Therefore, to numerically evaluate  the energy density (\ref{actionS55}), for the fermionic fields Eqs. (\ref{psi_expansion2}), with the normalizable profile functions (\ref{psi_expansion1}) and \eqref{fL_fR}, were used, whereas for the  part involving the gauge field strength, Eq. (\ref{VInt_q}), with the Fourier prescription (\ref{V_Fourier}), was employed. }

Therefore one can compute its Fourier transform and, subsequently, the modal fraction, and the DCE, respectively using Eqs. (\ref{fou}) -- (\ref{confige}).

The Lagrangian densities
are functions of the mixture parameters $d_3^{\scalebox{.6}{\textsc{$N^*N$}}}, d_4^{\scalebox{.6}{\textsc{$N^*N$}}}$, $d_5^{\scalebox{.6}{\textsc{$N^*N$}}}$, in (\ref{par2}), as well as the parameters $c_4^{\scalebox{.6}{\textsc{$N^*N$}}}$ and $c_5^{\scalebox{.6}{\textsc{$N^*N$}}}$ in (\ref{par1}). Their values in (\ref{par1}, \ref{par2}) were derived from a fit to the helicity amplitudes of the $\gamma N \to N^*(1535)$ transition \cite{Gutsche:2019yoo}. Instead, we can derive these parameters from first principles regarding information entropy, using the DCE. The strategy hereon is to take, in each one of the following analyses, three fixed parameters among the ones in Eqs. (\ref{par1}) and (\ref{par2}), leaving two free adjustable parameters to be determined out of the global minima of the DCE. As the nuclear physical system attains microstates of maximum stability, at the global minimum of the DCE, two among the mixture parameters, a priori free, will be derived, corresponding to the preferred dominant state occupied by the nuclear physical system, with maximal configurational stability. The lower the DCE is, the more localized the energy density is and the lower the uncertainty involved. Also, the higher the DCE, the higher the accuracy in determining the localization of the nuclear system, corresponding to the prevalent wave state occupied in momentum space.

\subsection{Deriving the mixture parameters $c_4^{\scalebox{.6}{\textsc{$N^*N$}}}$ and $c_5^{\scalebox{.6}{\textsc{$N^*N$}}}$}

\begin{figure}[H]
 \centering
 \includegraphics[width=4.5in]{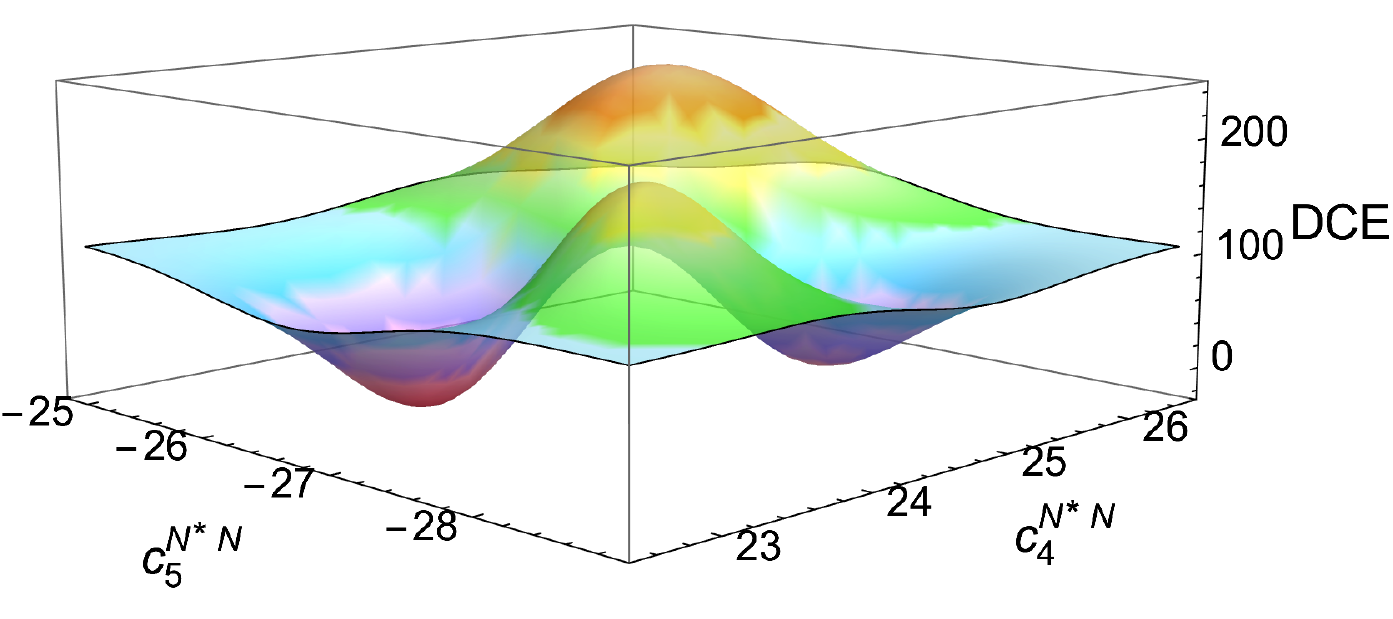}
 \caption{DCE as a function of $c_4^{\scalebox{.6}{\textsc{$N^*N$}}}$ and $c_5^{\scalebox{.6}{\textsc{$N^*N$}}}$. There is a global minimum DCE$_{\scalebox{0.6}{$\textsc{min}$}}$($c_4^{\scalebox{.6}{\textsc{$N^*N$}}}, c_5^{\scalebox{.6}{\textsc{$N^*N$}}}$) = 2.728 nat, for $c_4^{\scalebox{.6}{\textsc{$N^*N$}}}=25.401$ and $c_5^{\scalebox{.6}{\textsc{$N^*N$}}} = -26.244$.}
 \label{fff1}
\end{figure}

The first analysis comprises taking the parameters $d_3^{\scalebox{.6}{\textsc{$N^*N$}}}, d_4^{\scalebox{.6}{\textsc{$N^*N$}}}$, and $d_5^{\scalebox{.6}{\textsc{$N^*N$}}}$, fixed as in (\ref{par2}), and deriving the adjustable parameters $c_4^{\scalebox{.6}{\textsc{$N^*N$}}}$ and $c_5^{\scalebox{.6}{\textsc{$N^*N$}}}$, taken as free parameters. 
The global minimum of the DCE computed as a function of $c_4^{\scalebox{.6}{\textsc{$N^*N$}}}$ and $c_5^{\scalebox{.6}{\textsc{$N^*N$}}}$, can select the best choice of values for these parameters, corresponding to the most dominant state occupied by the nuclear quantum system. The DCE, employing Eqs. (\ref{fou}) -- (\ref{confige}), as a function of $c_4^{\scalebox{.6}{\textsc{$N^*N$}}}$ and $c_5^{\scalebox{.6}{\textsc{$N^*N$}}}$, is plotted in Fig. \ref{fff1}.
\noindent The global minimum (\textsc{gmin})
\beq
\text{DCE$\left(c_{4\,{\scalebox{0.6}{$\textsc{gmin}$}}}^{\scalebox{.6}{\textsc{$N^*N$}}}, c_{5\,{\scalebox{0.6}{$\textsc{gmin}$}}}^{\scalebox{.6}{\textsc{$N^*N$}}}\right)$ = 2.728 nat},\label{dce1}
\eeq
at
\beq
 \text{$c_{4\,{\scalebox{0.6}{$\textsc{gmin}$}}}^{\scalebox{.6}{\textsc{$N^*N$}}}=25.401$, \qquad\qquad $c_{5\,{\scalebox{0.6}{$\textsc{gmin}$}}}^{\scalebox{.6}{\textsc{$N^*N$}}} = -26.244$},\label{dce11}
 \eeq
matches data from the AdS/QCD soft-wall model in Ref. \cite{Gutsche:2019yoo} within an accuracy of 0.57\% and 2.43\%, respectively. The global minimum was derived through the routine {\tt NMinimize} in {\tt Mathematica} 13.0.0.0. The range of the ($c_4^{\scalebox{.6}{\textsc{$N^*N$}}}, c_5^{\scalebox{.6}{\textsc{$N^*N$}}}$)-parameter space here analyzed suffices to cover all the possibilities that are compatible to experimental nucleon masses and the Roper mass. In addition, the nuclear system here studied has higher configurational stability at the global minimum, being this mode more prevalent and dominant among all possible quantum states defined in the ($c_4^{\scalebox{.6}{\textsc{$N^*N$}}}, c_5^{\scalebox{.6}{\textsc{$N^*N$}}}$)-parameter space. In fact, the global minimum DCE$_{\scalebox{0.6}{$\textsc{min}$}}$($c_4^{\scalebox{.6}{\textsc{$N^*N$}}}, c_5^{\scalebox{.6}{\textsc{$N^*N$}}}$) = 2.728 nat, at the point $c_{4\,{\scalebox{0.6}{$\textsc{gmin}$}}}^{\scalebox{.6}{\textsc{$N^*N$}}}=25.401$ and $c_{5\,{\scalebox{0.6}{$\textsc{gmin}$}}}^{\scalebox{.6}{\textsc{$N^*N$}}} = -26.244$, corroborates respectively within an accuracy of 0.57\% and 2.43\%, comparing to data from the AdS/QCD soft-wall model in Ref. \cite{Gutsche:2019yoo}. This result is far from a trivial one.
Indeed, $c_4^{\scalebox{.6}{\textsc{$N^*N$}}}$ and $c_5^{\scalebox{.6}{\textsc{$N^*N$}}}$ were taken as completely free parameters to compute the DCE depicted in Fig. \ref{fff1}. Nothing beacons for specific values of these parameters but the fact that the global minimum of the DCE corresponds to the state attained by the nuclear system for which the configurational stability is maximal. Also, the global minimum of the DCE indicates that the nuclear system presents a higher data compression rate of information into the wave modes describing the spatial complexity of the system.

\begin{figure}[H]
 \centering
 \includegraphics[width=3.1in]{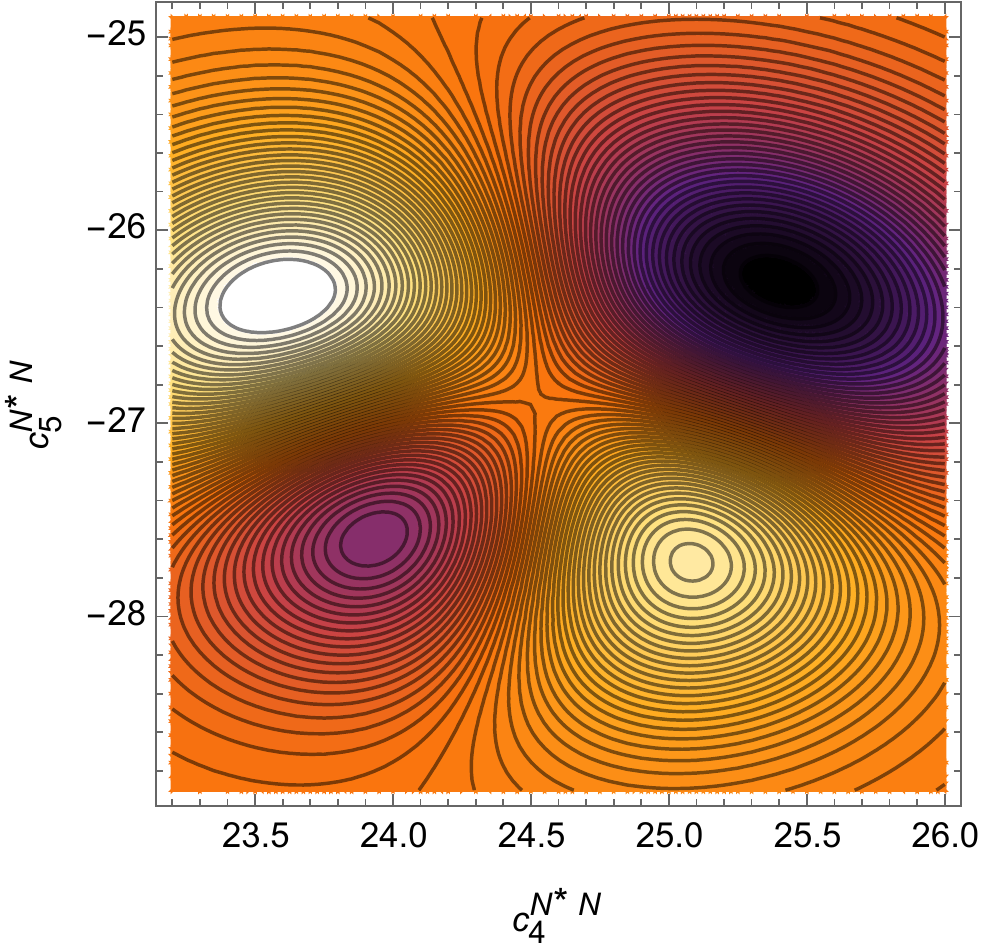}
 \caption{Contour plot of the
 DCE as a function of the parameters $c_4^{\scalebox{.6}{\textsc{$N^*N$}}}$ and $c_5^{\scalebox{.6}{\textsc{$N^*N$}}}$. }
 \label{fff2}
\end{figure}

Numerical analysis reveals that beyond the range in the plot in Fig. \ref{fff1} and \ref{fff2}, there is another local minimum (\textsc{lmin}),
\beq
\text{DCE$\left(c_{4\,{\scalebox{0.6}{$\textsc{lmin}$}}}^{\scalebox{.6}{\textsc{$N^*N$}}}, c_{5\,{\scalebox{0.6}{$\textsc{lmin}$}}}^{\scalebox{.6}{\textsc{$N^*N$}}}\right)$ = 23.431 nat},\label{dce2}
\eeq
at
\beq
 \text{$c_{4\,{\scalebox{0.6}{$\textsc{lmin}$}}}^{\scalebox{.6}{\textsc{$N^*N$}}}=23.838$,\qquad\qquad\qquad $c_{5\,{\scalebox{0.6}{$\textsc{lmin}$}}}^{\scalebox{.6}{\textsc{$N^*N$}}}= -27.582$.}\label{dce22}\eeq
 The local minimum
of the DCE \eqref{dce2} at the point \eqref{dce22} in the parameter space is the second most stable point of the system, however, the value of the DCE \eqref{dce2} is 859\% higher than the value of the DCE \eqref{dce1} at the global minimum. There are also two maxima, as seen in Fig. \ref{fff1}. They represent the values of
\beq
\text{DCE$\left(c_{4\,{\scalebox{0.6}{$\textsc{max1}$}}}^{\scalebox{.6}{\textsc{$N^*N$}}}, c_{5\,{\scalebox{0.6}{$\textsc{max1}$}}}^{\scalebox{.6}{\textsc{$N^*N$}}}\right)$ = 245.740 nat},\label{dce3}
\eeq
at the global maximum
\beq
 \text{$c_{4\,{\scalebox{0.6}{$\textsc{max1}$}}}^{\scalebox{.6}{\textsc{$N^*N$}}}=23.558$,\qquad\qquad $c_{5\,{\scalebox{0.6}{$\textsc{max1}$}}}^{\scalebox{.6}{\textsc{$N^*N$}}} = -26.381$,}\label{dce33}\eeq
 and
 \beq
\text{DCE$\left(c_{4\,{\scalebox{0.6}{$\textsc{max2}$}}}^{\scalebox{.6}{\textsc{$N^*N$}}}, c_{5\,{\scalebox{0.6}{$\textsc{max2}$}}}^{\scalebox{.6}{\textsc{$N^*N$}}}\right)$ = 171.533 nat},\label{dce4}
\eeq
at the local maximum
\beq
 \text{$c_{4\,{\scalebox{0.6}{$\textsc{max2}$}}}^{\scalebox{.6}{\textsc{$N^*N$}}}=25.089$,\qquad\qquad $c_{5\,{\scalebox{0.6}{$\textsc{max2}$}}}^{\scalebox{.6}{\textsc{$N^*N$}}} = -27.735$.}\label{dce44}\eeq
 wherein the nuclear system is most unstable, from the configurational point of view.

 The contour plot in Fig. \ref{fff2} exhibits the DCE as a function of parameters $c_4^{\scalebox{.6}{\textsc{$N^*N$}}}$ and $c_5^{\scalebox{.6}{\textsc{$N^*N$}}}$.
\noindent Regarding Fig. \ref{fff2}, the dark center of the purple bundle of closed curves, in the first quadrant, indicates the point (\ref{dce11}) corresponding to the global minimum of the DCE (\ref{dce1}). The center of the white region in the second quadrant corresponds to the DCE global maximum \eqref{dce3} at the point \eqref{dce33}, whereas the light-yellow center of the most inner ellipsis-like curve in the fourth quadrant refers to the local minimum \eqref{dce4} at the point \eqref{dce44}. Besides, the purple center of the most inner ellipsis-like curve in the third quadrant refers to the local minimum \eqref{dce22} at the point \eqref{dce2}.

The $(c_4^{\scalebox{.6}{\textsc{$N^*N$}}}, c_5^{\scalebox{.6}{\textsc{$N^*N$}}})$-parameter space undergoes a
 splitting into configurational isentropic subsectors. The contour plot represents curves joining points of the parameter space attaining an equal value of the DCE. The gradient of the DCE is orthogonal to the contour lines corresponding to configurational isentropic curves. One can compare Figs. \ref{fff1} and \ref{fff2}, realizing that when the contour lines are close together, the variation of the DCE is steeper with respect to the parameters $c_4^{\scalebox{.6}{\textsc{$N^*N$}}}$ and $c_5^{\scalebox{.6}{\textsc{$N^*N$}}}$. The outer [inner] isentropic subsectors, with hotter [colder] colors, regard higher [lower] values of the DCE. The dark purple subsector encircles the global minimum DCE${}_{\scalebox{0.6}{$\textsc{gmin}$}}$($c_4^{\scalebox{.6}{\textsc{$N^*N$}}},c_5^{\scalebox{.6}{\textsc{$N^*N$}}})$.

\subsection{Deriving the mixture parameters $d_4^{\scalebox{.6}{\textsc{$N^*N$}}}$ and $d_5^{\scalebox{.6}{\textsc{$N^*N$}}}$}

\begin{figure}[H]
 \centering
 \includegraphics[width=5.2in]{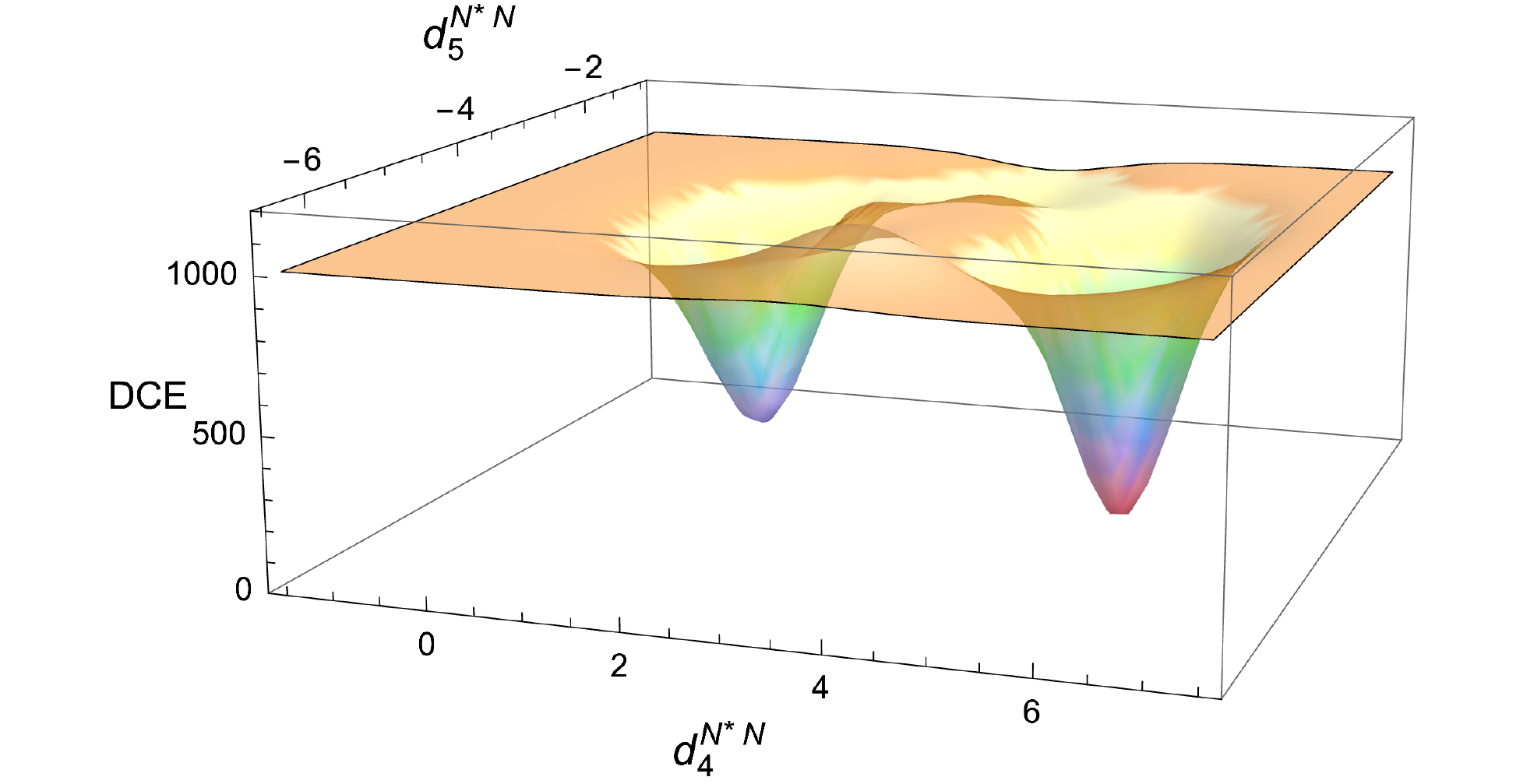}
 \caption{DCE as a function of $d_4^{\scalebox{.6}{\textsc{$N^*N$}}}$ and $d_5^{\scalebox{.6}{\textsc{$N^*N$}}}$. There is a global minimum DCE$_{\scalebox{0.6}{$\textsc{min}$}}$($d_4^{\scalebox{.6}{\textsc{$N^*N$}}}, d_5^{\scalebox{.6}{\textsc{$N^*N$}}}$) = 98.733 nat, for $d_4^{\scalebox{.6}{\textsc{$N^*N$}}}=5.634$ and $d_5^{\scalebox{.6}{\textsc{$N^*N$}}} = -3.531$, respectively within an accuracy of 0.12\% and 1.37\%, comparing to data from the AdS/QCD soft-wall model in Ref. \cite{Gutsche:2019yoo}.}
 \label{fff3}
\end{figure}

\begin{figure}[H]
 \centering
 \includegraphics[width=3.1in]{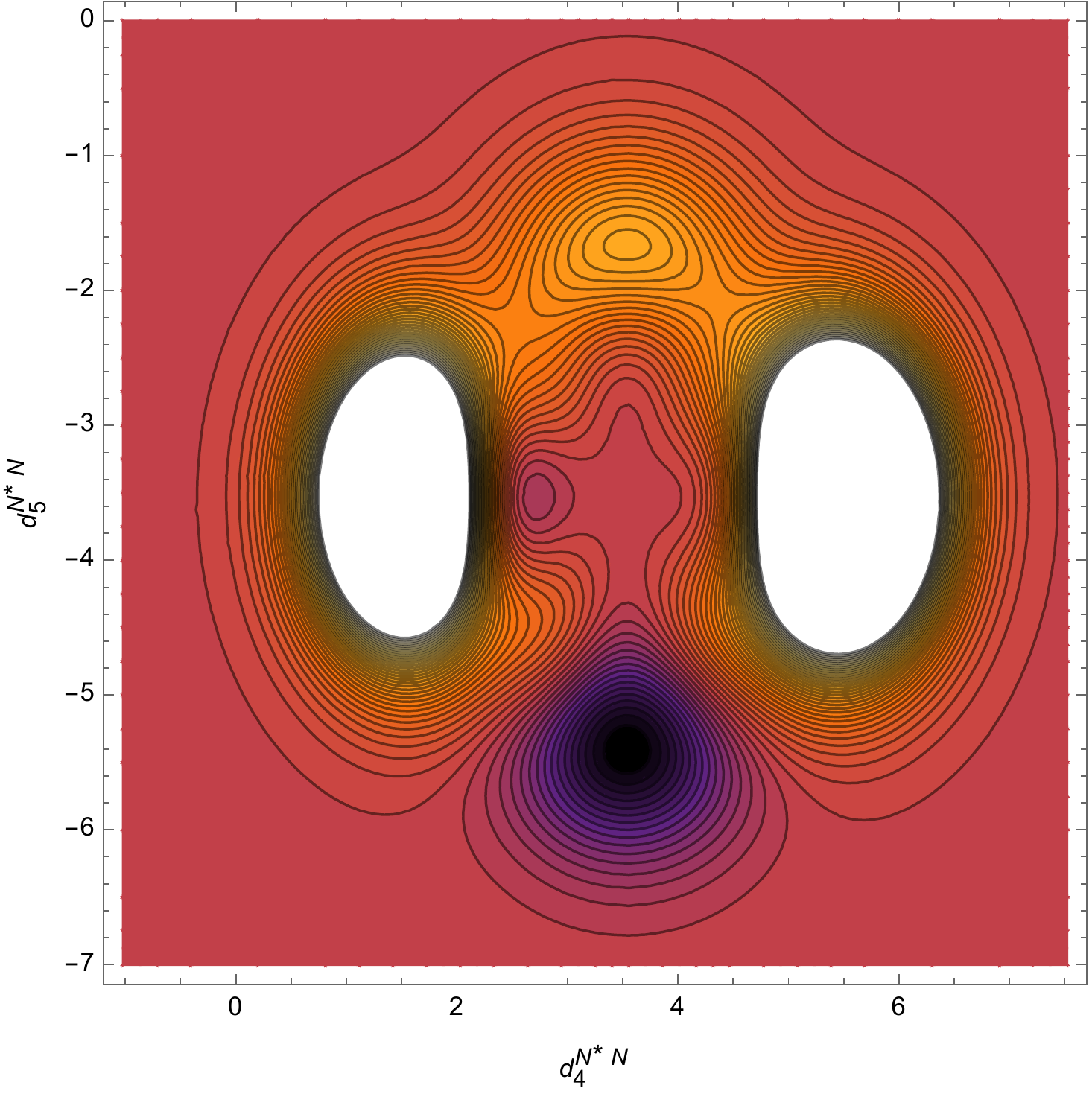}
 \caption{Contour plot of the
 DCE as a function of the parameters $d_4^{\scalebox{.6}{\textsc{$N^*N$}}}$ and $d_5^{\scalebox{.6}{\textsc{$N^*N$}}}$. }
 \label{fff4}
\end{figure}

Now, a second analysis regards assuming the parameters $d_3^{\scalebox{.6}{\textsc{$N^*N$}}}, c_4^{\scalebox{.6}{\textsc{$N^*N$}}}$, and $c_5^{\scalebox{.6}{\textsc{$N^*N$}}}$ in (\ref{par2}) and letting free the adjustable mixture parameters $d_4^{\scalebox{.6}{\textsc{$N^*N$}}}$ and $d_5^{\scalebox{.6}{\textsc{$N^*N$}}}$. Once the DCE is computed as a function of. $d_4^{\scalebox{.6}{\textsc{$N^*N$}}}$ and $d_5^{\scalebox{.6}{\textsc{$N^*N$}}}$, the DCE global minimum will indicate the
preferred values of $d_4^{\scalebox{.6}{\textsc{$N^*N$}}}$ and $d_5^{\scalebox{.6}{\textsc{$N^*N$}}}$ which correspond to the point of higher configurational stability occupied by the nuclear system. We will show that the respective values of the mixture parameters $d_4^{\scalebox{.6}{\textsc{$N^*N$}}}$ and $d_5^{\scalebox{.6}{\textsc{$N^*N$}}}$ corroborate to phenomenological data with great accuracy. The DCE can be then computed, using the protocol Eqs. (\ref{fou}) -- (\ref{confige}), as a function of $d_4^{\scalebox{.6}{\textsc{$N^*N$}}}$ and $d_5^{\scalebox{.6}{\textsc{$N^*N$}}}$, is plotted in Fig. \ref{fff3}.
\noindent The global minimum (\textsc{gmin})
\beq
\text{DCE$\left(d_{4\,{\scalebox{0.6}{$\textsc{gmin}$}}}^{\scalebox{.6}{\textsc{$N^*N$}}}, d_{5\,{\scalebox{0.6}{$\textsc{gmin}$}}}^{\scalebox{.6}{\textsc{$N^*N$}}}\right)$ = 98.733 nat},\label{dce12a}
\eeq
at
\beq
 \text{$d_{4\,{\scalebox{0.6}{$\textsc{gmin}$}}}^{\scalebox{.6}{\textsc{$N^*N$}}}=5.634$,\qquad\qquad $d_{5\,{\scalebox{0.6}{$\textsc{gmin}$}}}^{\scalebox{.6}{\textsc{$N^*N$}}} = -3.531$},\label{dce112}
 \eeq
matches data from the AdS/QCD soft-wall model in Ref. \cite{Gutsche:2019yoo} within an accuracy of 0.12\% and 1.37\%, respectively. In addition, the nucleonic transition has higher configurational stability at the global minimum, which is a dominant state also from phenomenological aspects.
Numerical analysis reveals that beyond the range in the plot in Fig. \ref{fff3} and \ref{fff4}, there are other two local minima (\textsc{lmin}).
The first one has the associated DCE given by
\beq
\text{DCE$\left(d_{4\,{\scalebox{0.6}{$\textsc{lmin1}$}}}^{\scalebox{.6}{\textsc{$N^*N$}}}, d_{5\,{\scalebox{0.6}{$\textsc{lmin1}$}}}^{\scalebox{.6}{\textsc{$N^*N$}}}\right)$ = 227.009 nat},\label{dce2a}
\eeq
at the point of the $\left(d_{4}^{\scalebox{.6}{\textsc{$N^*N$}}}, d_{5}^{\scalebox{.6}{\textsc{$N^*N$}}}\right)$-parameter space
\beq
 \text{$d_{4\,{\scalebox{0.6}{$\textsc{lmin1}$}}}^{\scalebox{.6}{\textsc{$N^*N$}}}=1.497$,\qquad\qquad\qquad $d_{5\,{\scalebox{0.6}{$\textsc{lmin1}$}}}^{\scalebox{.6}{\textsc{$N^*N$}}}= -3.508$.}\label{dce22a}\eeq
 The first local minimum
of the DCE at the parameter space is the second most stable point of the system, however, the value of the DCE \eqref{dce2a} is 129.92\% higher than the value of the DCE \eqref{dce12a} at the global minimum. Besides, the second local minimum has associated DCE given by
\beq
\text{DCE$\left(d_{4\,{\scalebox{0.6}{$\textsc{lmin2}$}}}^{\scalebox{.6}{\textsc{$N^*N$}}}, d_{5\,{\scalebox{0.6}{$\textsc{lmin2}$}}}^{\scalebox{.6}{\textsc{$N^*N$}}}\right)$ = 890.539 nat},\label{dce222a}
\eeq
at the point of the parameter space
\beq
 \text{$d_{4\,{\scalebox{0.6}{$\textsc{lmin2}$}}}^{\scalebox{.6}{\textsc{$N^*N$}}}=3.589$,\qquad\qquad\qquad $d_{5\,{\scalebox{0.6}{$\textsc{lmin2}$}}}^{\scalebox{.6}{\textsc{$N^*N$}}}= -1.677$.}\label{dce22a}\eeq
 The second local minimum
of the DCE at the parameter space is the second most stable point of the system, however, the value of the DCE \eqref{dce222a} is 801.96\% higher than the value of the DCE \eqref{dce12a} at the global minimum.

There are also two maxima, as seen in Fig. \ref{fff3}. They represent the values of
\beq
\text{DCE$\left(d_{4\,{\scalebox{0.6}{$\textsc{max1}$}}}^{\scalebox{.6}{\textsc{$N^*N$}}}, d_{5\,{\scalebox{0.6}{$\textsc{max1}$}}}^{\scalebox{.6}{\textsc{$N^*N$}}}\right)$ = 1167.845 nat},\label{dce3a}
\eeq
at the global maximum
\beq
 \text{$d_{4\,{\scalebox{0.6}{$\textsc{max1}$}}}^{\scalebox{.6}{\textsc{$N^*N$}}}=3.237$,\qquad\qquad $d_{5\,{\scalebox{0.6}{$\textsc{max1}$}}}^{\scalebox{.6}{\textsc{$N^*N$}}} = -5.412$,}\label{dce33a}\eeq
 and, for the local maximum,
 \beq
\text{DCE$\left(d_{4\,{\scalebox{0.6}{$\textsc{max2}$}}}^{\scalebox{.6}{\textsc{$N^*N$}}}, d_{5\,{\scalebox{0.6}{$\textsc{max2}$}}}^{\scalebox{.6}{\textsc{$N^*N$}}}\right)$ = 1045.041 nat},\label{dce4a}
\eeq
at \beq
 \text{$d_{4\,{\scalebox{0.6}{$\textsc{max2}$}}}^{\scalebox{.6}{\textsc{$N^*N$}}}=2.730$,\qquad\qquad $d_{5\,{\scalebox{0.6}{$\textsc{max2}$}}}^{\scalebox{.6}{\textsc{$N^*N$}}} = -3.599$.}\label{dce44a}\eeq
 wherein the nuclear system presents a higher configurational instability. 

 The contour plot in Fig. \ref{fff4} exhibits the DCE as a function of parameters $d_4^{\scalebox{.6}{\textsc{$N^*N$}}}$ and $d_5^{\scalebox{.6}{\textsc{$N^*N$}}}$.
 \noindent Fig. \ref{fff4} illustrates the center of the white inner region at the right corresponding to the global minimum of the DCE (\ref{dce12a}).
 The local minimum of the DCE, \eqref{dce2a}, at the point \eqref{dce22a}, is illustrated by the center of the white inner region at the left. The second local minimum of the DCE, \eqref{dce222a}, at the point of the parameter space
\eqref{dce22a}, is located at the center of the orange elliptic region inner region. Besides, the global maximum of the DCE, \eqref{dce3a},
at the point \eqref{dce33a}, is represented by the dark center of the purple bundle in the south region of Fig. \ref{fff4}, while the local maximum of the DCE \eqref{dce4a} is located at the point \eqref{dce44a} in the
$\left(d_{4\,{\scalebox{0.6}{$\textsc{max2}$}}}^{\scalebox{.6}{\textsc{$N^*N$}}}, d_{5\,{\scalebox{0.6}{$\textsc{max2}$}}}^{\scalebox{.6}{\textsc{$N^*N$}}}\right)$-parameter space depicted by the purple region slightly dislocated off the center in Fig. \ref{fff4}. One can realize that the $(d_4^{\scalebox{.6}{\textsc{$N^*N$}}}, d_5^{\scalebox{.6}{\textsc{$N^*N$}}})$-parameter space is also split into differential configurational isentropic subsectors separated by inhomogeneous gaps, with a variation $\Delta [{\rm DCE}(d_4^{\scalebox{.6}{\textsc{$N^*N$}}}, d_5^{\scalebox{.6}{\textsc{$N^*N$}}})] \sim 3$, in average.

\subsection{Deriving the mixture parameters $d_3^{\scalebox{.6}{\textsc{$N^*N$}}}$ and $d_4^{\scalebox{.6}{\textsc{$N^*N$}}}$}

\begin{figure}[H]
 \centering
 \includegraphics[width=4.99in]{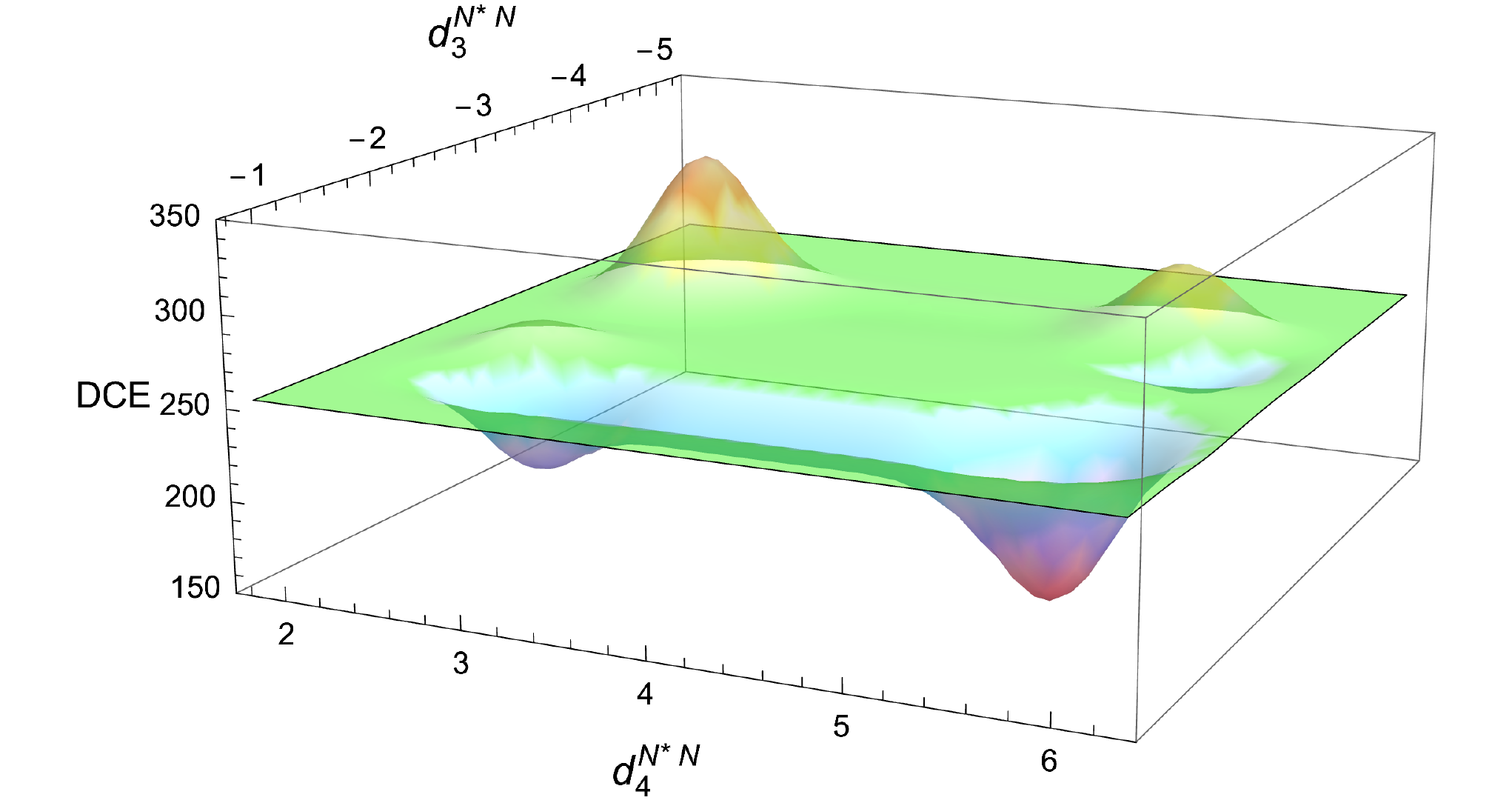}
 \caption{DCE as a function of $d_3^{\scalebox{.6}{\textsc{$N^*N$}}}$ and $d_4^{\scalebox{.6}{\textsc{$N^*N$}}}$. There is a global minimum DCE$_{\scalebox{0.6}{$\textsc{min}$}}$($d_{3\,{\scalebox{0.6}{$\textsc{gmin}$}}}^{\scalebox{.6}{\textsc{$N^*N$}}}, d_{4\,{\scalebox{0.6}{$\textsc{gmin}$}}}^{\scalebox{.6}{\textsc{$N^*N$}}}$) = 168.755 nat, for $d_{3\,{\scalebox{0.6}{$\textsc{gmin}$}}}^{\scalebox{.6}{\textsc{$N^*N$}}}=-1.857$ and $d_{4\,{\scalebox{0.6}{$\textsc{gmin}$}}}^{\scalebox{.6}{\textsc{$N^*N$}}} = 5.538$, respectively within an accuracy of 1.74\% and 2.01\%, comparing to data from the AdS/QCD soft-wall model in Ref. \cite{Gutsche:2019yoo}.}
 \label{fff5}
\end{figure}

The third way to analyze the mixture parameters constituting
the linear combination that defines the interaction Lagrangian density (\ref{actionS4}), we assume the values of the parameters $c_4^{\scalebox{.6}{\textsc{$N^*N$}}}, c_5^{\scalebox{.6}{\textsc{$N^*N$}}}$, and $d_5^{\scalebox{.6}{\textsc{$N^*N$}}}$ in (\ref{par2}) and letting free the adjustable mixture parameters $d_3^{\scalebox{.6}{\textsc{$N^*N$}}}$ and $d_4^{\scalebox{.6}{\textsc{$N^*N$}}}$. Subsequently computing the DCE as a scalar function of $d_3^{\scalebox{.6}{\textsc{$N^*N$}}}$ and $d_4^{\scalebox{.6}{\textsc{$N^*N$}}}$, its global minimum will point to the
preferred values of the parameters $d_3^{\scalebox{.6}{\textsc{$N^*N$}}}$ and $d_4^{\scalebox{.6}{\textsc{$N^*N$}}}$ representing the point of higher configurational stability occupied by the nuclear system, also complying with phenomenological data with good accuracy. Therefore the DCE can be computed, employing the procedure dictated by Eqs. (\ref{fou}) -- (\ref{confige}), as a function of $d_3^{\scalebox{.6}{\textsc{$N^*N$}}}$ and $d_4^{\scalebox{.6}{\textsc{$N^*N$}}}$. The DCE is plotted in Fig. \ref{fff5}.
\noindent The global minimum (\textsc{gmin})
\beq
\text{DCE$_{\scalebox{0.6}{$\textsc{min}$}}$($d_{3\,{\scalebox{0.6}{$\textsc{gmin}$}}}^{\scalebox{.6}{\textsc{$N^*N$}}}, d_{4\,{\scalebox{0.6}{$\textsc{gmin}$}}}^{\scalebox{.6}{\textsc{$N^*N$}}}$) = 168.755 nat},\label{dce13}
\eeq
at
\beq
 \text{$d_{3\,{\scalebox{0.6}{$\textsc{gmin}$}}}^{\scalebox{.6}{\textsc{$N^*N$}}}=-1.857$ and $d_{4\,{\scalebox{0.6}{$\textsc{gmin}$}}}^{\scalebox{.6}{\textsc{$N^*N$}}} = 5.538$},\label{dce113}
 \eeq
matches data from the AdS/QCD soft-wall model in Ref. \cite{Gutsche:2019yoo} within an accuracy of 1.74\% and 2.01\%, respectively. The range of the ($d_3^{\scalebox{.6}{\textsc{$N^*N$}}}, d_4^{\scalebox{.6}{\textsc{$N^*N$}}}$)-parameter space here analyzed suffices to cover all the experimental possibilities. In addition, the nuclear system here studied has higher configurational stability at the global minimum, being this mode more prevalent and dominant also from the experimental point of view. In fact, the global minimum DCE$_{\scalebox{0.6}{$\textsc{min}$}}$($d_{3\,{\scalebox{0.6}{$\textsc{gmin}$}}}^{\scalebox{.6}{\textsc{$N^*N$}}}, d_{4\,{\scalebox{0.6}{$\textsc{gmin}$}}}^{\scalebox{.6}{\textsc{$N^*N$}}}$) = 168.755 nat, for $d_{3\,{\scalebox{0.6}{$\textsc{gmin}$}}}^{\scalebox{.6}{\textsc{$N^*N$}}}=-1.857$ and $d_{4\,{\scalebox{0.6}{$\textsc{gmin}$}}}^{\scalebox{.6}{\textsc{$N^*N$}}} = 5.538$, corroborates respectively within an accuracy of 1.74\% and 2.01\%, comparing to data from the AdS/QCD soft-wall model in Ref. \cite{Gutsche:2019yoo}.
Indeed, the parameters $d_3^{\scalebox{.6}{\textsc{$N^*N$}}}$ and $d_4^{\scalebox{.6}{\textsc{$N^*N$}}}$ are, a priori, free parameters. Hence, the global minimum of the DCE, corresponding to the state occupied by the nucleonic system has maximal configurational stability. Also, the information underlying the nuclear system presents a higher data compression rate into the wave modes describing the spatial complexity of the system.

Numerical analysis reveals that beyond the range in the plot in Fig. \ref{fff5}, there are other local minima (\textsc{lmin}). For the first one,
\beq
\text{DCE$\left(d_{3\,{\scalebox{0.6}{$\textsc{lmin1}$}}}^{\scalebox{.6}{\textsc{$N^*N$}}}, d_{4\,{\scalebox{0.6}{$\textsc{lmin1}$}}}^{\scalebox{.6}{\textsc{$N^*N$}}}\right)$ = 226.874 nat},\label{dce230}
\eeq
at
\beq
 \text{$d_{3\,{\scalebox{0.6}{$\textsc{lmin1}$}}}^{\scalebox{.6}{\textsc{$N^*N$}}}=-1.719$,\qquad\qquad\qquad $d_{4\,{\scalebox{0.6}{$\textsc{lmin}$}}}^{\scalebox{.6}{\textsc{$N^*N$}}}= 2.908$}\label{dce223},\eeq
 and
 \beq
\text{DCE$\left(d_{3\,{\scalebox{0.6}{$\textsc{lmin2}$}}}^{\scalebox{.6}{\textsc{$N^*N$}}}, d_{4\,{\scalebox{0.6}{$\textsc{lmin2}$}}}^{\scalebox{.6}{\textsc{$N^*N$}}}\right)$ = 5.780 nat},\label{dce231}
\eeq
at
\beq
 \text{$d_{3\,{\scalebox{0.6}{$\textsc{lmin}$}}}^{\scalebox{.6}{\textsc{$N^*N$}}}=-3.321$,\qquad\qquad\qquad $d_{4\,{\scalebox{0.6}{$\textsc{lmin}$}}}^{\scalebox{.6}{\textsc{$N^*N$}}}= 5.769$.}\label{dce2231}\eeq

There are also three maxima, as seen in Fig. \ref{fff5}. They represent the values of
\beq
\text{DCE$\left(d_{3\,{\scalebox{0.6}{$\textsc{max1}$}}}^{\scalebox{.6}{\textsc{$N^*N$}}}, d_{4\,{\scalebox{0.6}{$\textsc{max1}$}}}^{\scalebox{.6}{\textsc{$N^*N$}}}\right)$ = 280.473 nat},\label{dce3aa}
\eeq
at the first local maximum (\textsc{max1})
\beq
 \text{$d_{3\,{\scalebox{0.6}{$\textsc{max1}$}}}^{\scalebox{.6}{\textsc{$N^*N$}}}=-3.860$,\qquad\qquad $d_{4\,{\scalebox{0.6}{$\textsc{max1}$}}}^{\scalebox{.6}{\textsc{$N^*N$}}} = 5.416$,}\label{dce33b}\eeq
the second local maximum (\textsc{max2})
\beq
\text{DCE$\left(d_{3\,{\scalebox{0.6}{$\textsc{max2}$}}}^{\scalebox{.6}{\textsc{$N^*N$}}}, d_{4\,{\scalebox{0.6}{$\textsc{max2}$}}}^{\scalebox{.6}{\textsc{$N^*N$}}}\right)$ = 261.381 nat},\label{dce3d1}
\eeq
at
\beq
 \text{$d_{3\,{\scalebox{0.6}{$\textsc{max1}$}}}^{\scalebox{.6}{\textsc{$N^*N$}}}=-2.228$,\qquad\qquad $d_{4\,{\scalebox{0.6}{$\textsc{max1}$}}}^{\scalebox{.6}{\textsc{$N^*N$}}} = 2.578$,}\label{dce33d}\eeq
whereas the global maximum (\textsc{gmax}) has differential configurational entropy given by
\beq
\text{DCE$\left(d_{3\,{\scalebox{0.6}{$\textsc{gmax}$}}}^{\scalebox{.6}{\textsc{$N^*N$}}}, d_{4\,{\scalebox{0.6}{$\textsc{gmax}$}}}^{\scalebox{.6}{\textsc{$N^*N$}}}\right)$ = 339.581 nat},\label{dce3ab}
\eeq
at the parameter space point
\beq
 \text{$d_{3\,{\scalebox{0.6}{$\textsc{gmax}$}}}^{\scalebox{.6}{\textsc{$N^*N$}}}=-3.713$,\qquad\qquad $d_{4\,{\scalebox{0.6}{$\textsc{gmax}$}}}^{\scalebox{.6}{\textsc{$N^*N$}}} = 2.675$,}\label{dce33ab}\eeq wherein the nuclear system is most unstable, from the configurational point of view.

\begin{figure}[H]
 \centering
 \includegraphics[width=3.1in]{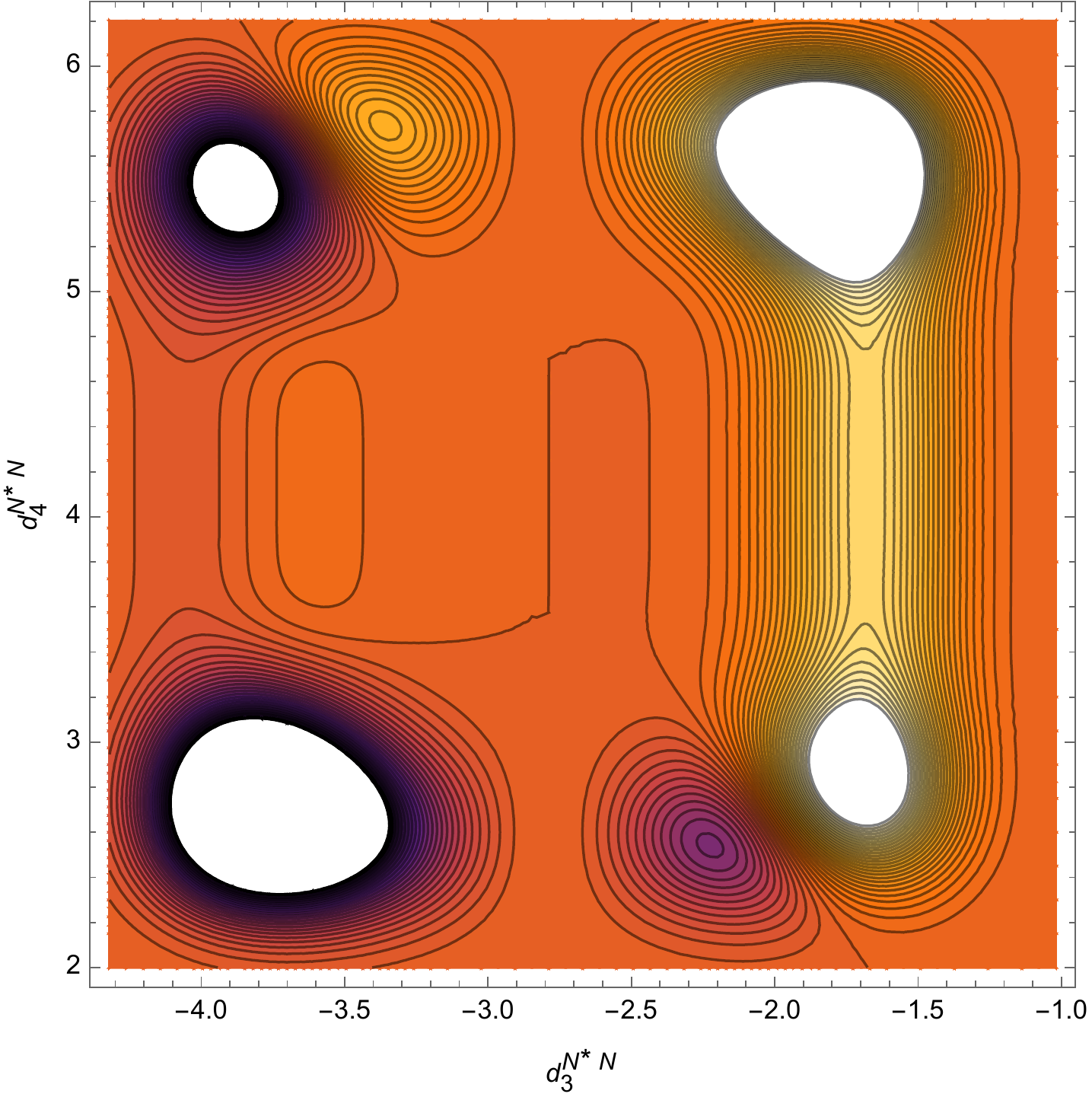}
 \caption{Contour plot of the
 DCE as a function of the parameters $d_3^{\scalebox{.6}{\textsc{$N^*N$}}}$ and $d_4^{\scalebox{.6}{\textsc{$N^*N$}}}$. }
 \label{fff6}
\end{figure}

 The contour plot in Fig. \ref{fff6} exhibits the DCE as a function of parameters $d_3^{\scalebox{.6}{\textsc{$N^*N$}}}$ and $d_4^{\scalebox{.6}{\textsc{$N^*N$}}}$.
\noindent Fig. \ref{fff6} explicitly shows the minima and maxima of the DCE, with respect to the parameters $d_3^{\scalebox{.6}{\textsc{$N^*N$}}}$ and $d_4^{\scalebox{.6}{\textsc{$N^*N$}}}$.
Besides, the $(d_3^{\scalebox{.6}{\textsc{$N^*N$}}}, d_4^{\scalebox{.6}{\textsc{$N^*N$}}})$-parameter space goes through a
 splitting into differential configurational isentropic subsectors.

Since the Fourier transform of the energy density encoded in Eq. (\ref{em1}) represents a random probability distribution, the DCE measures the shape of complexity underlying the nuclear system.
All the peaks and valleys of the DCE, respectively corresponding to maxima and minima in Figs. \ref{fff1}, \ref{fff3}, and \ref{fff5}, present kurtosis-like features.
The region surrounding the peaks and valleys of the DCE in Fig. \ref{fff1} approaches a mesokurtic profile, whereas the two valleys of the DCE in Fig. \ref{fff3} present a leptokurtic form. The DCE in Fig. \ref{fff5} shows a platykurtic shape in the neighborhood of the maxima and minima, producing less extreme outliers than the normal distribution if one interprets the DCE itself as a distribution dependent on two parameters.

\section{Conclusions}
\label{IV}
The soft-wall AdS/QCD model was used to study the $\gamma N \to N^*(1535)$ transition, including the coupling of two fermionic fields describing a nucleon and its respective resonance in the AdS bulk.
The DCE was shown to be a powerful tool for deriving adjustable mixture parameters composing the Lagrangian density that expresses interactions between nucleons, their respective resonances, and the gauge vector field in the $\gamma N \to N^*(1535)$ transition. The DCE that underlies this quantum transition was computed in three possible analyses.
Each analysis takes the DCE as a function of two among five adjustable parameters (\ref{par1}, \ref{par2}) that define the minimal and nonminimal couplings in the nuclear interaction (\ref{actionS4}). It shows that the DCE can drive the choice of
parameters entering several aspects of phenomenological bottom-up approaches of AdS/QCD, complying with experimental data with good accuracy. The critical points of the DCE match the corresponding points in the domain of the parameter space for which information in the nuclear system has a higher compression rate, with lower uncertainty and higher configurational stability. Therefore, the points in the domain of the parameter space
(\ref{dce11}), (\ref{dce112}), and (\ref{dce113}), at which the respective global minima of the DCE (\ref{dce1}), (\ref{dce12a}), (\ref{dce13}), are respectively attained, corroborate to the values (\ref{par1}) and (\ref{par2}), obtained from a suitable fit involving to the transversal and longitudinal helicity amplitudes of the $\gamma N \to N^*(1535)$ transition \cite{Gutsche:2019yoo}.
It points to the DCE setup, in the context of AdS/QCD, as a suitable technique to investigate nuclear interactions.
One can use this procedure to evaluate and predict feasible values for other free parameters that compose several phenomenological approaches in AdS/QCD.
In particular, as a feasible perspective, one can use the recent results in Ref. \cite{Ramalho:2017pyc} to emulate the results here obtained to deeper study the $\gamma^\ast N \to N(1440)$ transition. Also, other electromagnetic transitions involving nucleons and their respective higher-spin resonances can be investigated in this context.
\bigskip
\bigskip\bigskip\bigskip

\subsubsection*{Acknowledgments}

RdR~is grateful to The S\~ao Paulo Research Foundation -- FAPESP (Grants No. 2022/01734-7 and No. 2021/01089-1) and the National Council for Scientific and Technological Development -- CNPq Grant No. 303390/2019-0, for partial financial support.
\medbreak


\begin{thebibliography}{999}



\bibitem{Gleiser:2011di} M. Gleiser and N. Stamatopoulos, Phys.\ Lett.\ B {\bf 713} (2012) 304 [arXiv:1111.5597 [hep-th].

\bibitem{Gleiser:2018kbq} M. Gleiser, M. Stephens and D. Sowinski, Phys.\ Rev.\ D {\bf 97} (2018) 096007 [arXiv:1803.08550 [hep-th]].

\bibitem{Gleiser:2012tu} M. Gleiser and N. Stamatopoulos, Phys.\ Rev.\ D {\bf 86} (2012) 045004 [arXiv:1205.3061 [hep-th]].


\bibitem{Gleiser:2018jpd}
M.~Gleiser and D.~Sowinski,
Phys. Rev. D \textbf{98} (2018)  056026
[arXiv:1807.07588 [hep-th]].

\bibitem{Bernardini:2016hvx} A. E. Bernardini and R. da  Rocha, Phys.\ Lett.\ B {\bf 762} (2016) 107 [arXiv:1605.00294 [hep-th]].


\bibitem{Sowinski:2015cfa} M. Gleiser and D. Sowinski, Phys.\ Lett.\ B {\bf 747} (2015) 125 [arXiv:1501.06800 [hep-th]].







\bibitem{Karapetyan:2018yhm} G. Karapetyan, Phys.\ Lett.\ B {\bf 786} (2018) 418 [arXiv:1807.04540 [nucl-th]].

\bibitem{Karapetyan:epjp} G. Karapetyan, Eur. Phys. J. Plus {\bf 136} (2021) 1012 [arXiv:2105.07546 [hep-ph]].

\bibitem{Karapetyan:plb} G. Karapetyan, Phys.\ Lett.\ B {\bf 781}  (2018) 201 [arXiv:1802.09105 [hep-ph]].

\bibitem{Karapetyan:2020epl} G. Karapetyan, EPL {\bf 129}  (2020) 18002 [arXiv:1912.10071 [hep-ph]].


\bibitem{Ma:2018wtw}
	C.~W.~Ma and Y.~G.~Ma,
	Prog.\ Part.\ Nucl.\ Phys.\  {\bf 99} (2018) 120
	[arXiv:1801.02192 [nucl-th]].

\bibitem{Karapetyan:2019fst}
G.~Karapetyan,
EPL \textbf{125} (2019) 58001
[arXiv:1901.05349 [hep-ph]].

\bibitem{Aharony:1999ti}
O.~Aharony, S.~S.~Gubser, J.~M.~Maldacena, H.~Ooguri and Y.~Oz,
Phys. Rept. \textbf{323} (2000) 183
 [arXiv:hep-th/9905111 [hep-th]].

	\bibitem{Witten:1998qj}
	E.~Witten,
	Adv.\ Theor.\ Math.\ Phys.\  {\bf 2} (1998) 253
	[hep-th/9802150].

\bibitem{Brodsky:2010ur}
S.~J.~Brodsky, G.~F.~de Teramond and A.~Deur,
Phys. Rev. D \textbf{81} (2010) 096010
[arXiv:1002.3948 [hep-ph]].

\bibitem{Brodsky:2014yha} S.~J.~Brodsky, G.~F.~de Teramond, H.~G.~Dosch, J.~Erlich,
	Phys.\ Rept.\  {\bf 584} (2015) 1
	[{arXiv:1407.8131 [hep-ph]}].

\bibitem{Vega:2008te}
A.~Vega and I.~Schmidt,
Phys. Rev. D \textbf{79} (2009) 055003
[arXiv:0811.4638 [hep-ph]].

\bibitem{Karch:2006pv}
A.~Karch, E.~Katz, D.~T.~Son and M.~A.~Stephanov,
Phys. Rev. D \textbf{74} (2006) 015005
[arXiv:hep-ph/0602229 [hep-ph]].

\bibitem{Gutsche:2012bp}
T.~Gutsche, V.~E.~Lyubovitskij, I.~Schmidt and A.~Vega,
Phys. Rev. D \textbf{86} (2012) 036007
[arXiv:1204.6612 [hep-ph]].

\bibitem{Gutsche:2011vb}
T.~Gutsche, V.~E.~Lyubovitskij, I.~Schmidt and A.~Vega,
Phys. Rev. D \textbf{85} (2012) 076003
[arXiv:1108.0346 [hep-ph]].

\bibitem{daRocha:2021ntm}
R.~da Rocha,
Phys. Rev. D \textbf{103} (2021) 106027
[arXiv:2103.03924 [hep-ph]].





\bibitem{Ferreira:2020iry}
L.~F.~Ferreira and R.~da Rocha,
Phys. Rev. D \textbf{101} (2020) 106002
[arXiv:2004.04551 [hep-th]].



\bibitem{Bazeia:2021stz}
D.~Bazeia and E.~I.~B.~Rodrigues,
Phys. Lett. A \textbf{392} (2021) 127170.



\bibitem{Braga:2021fey}
N.~R.~F.~Braga, Y.~F.~Ferreira and L.~F.~Ferreira,
Phys. Rev. D \textbf{105} (2022) 114044
[arXiv:2110.04560 [hep-th]].


\bibitem{Braga:2021zyi}
N.~R.~F.~Braga and O.~C.~Junqueira,
Phys. Lett. B \textbf{820} (2021) 136485
[arXiv:2105.12347 [hep-th]].

\bibitem{Braga:2020hhs}
N.~R.~F.~Braga and R.~da Mata,
Phys. Lett. B \textbf{811} (2020) 135918
[arXiv:2008.10457 [hep-th]].

\bibitem{Bernardini:2018uuy} A. E. Bernardini and R. da Rocha, Phys.\ Rev.\ D {\bf 98} (2018) 126011 [arXiv:1809.10055 [hep-th]].

\bibitem{Braga:2018fyc} N. R. F. Braga, L. F. Ferreira and R. da  Rocha, Phys.\ Lett.\ B {\bf 787} (2018) 16 [arXiv:1808.10499 [hep-ph]].



\bibitem{Braga:2017fsb} N. R. F. Braga and R. da  Rocha, Phys.\ Lett.\ B {\bf 776} (2018) 78 [arXiv:1710.07383 [hep-th]].


\bibitem{Colangelo:2018mrt} P. Colangelo and F. Loparco, Phys.\ Lett.\ B {\bf 788} (2019) 500 [arXiv:1811.05272 [hep-ph]].

\bibitem{pdg} P. A. Zyla et al. (Particle Data Group), Prog. Theor. Exp. Phys. {\bf 2022} (2022) 083C01.




\bibitem{Karapetyan:2017edu} G. Karapetyan, EPL {\bf 118}  (2017) 38001 [arXiv:1705.1061 [hep-ph]].


\bibitem{Karapetyan:2016fai} G. Karapetyan, EPL {\bf 117}  (2017) 18001
[arXiv:1612.09564  [hep-ph]].

\bibitem{Karapetyan:2021crv}
G.~Karapetyan,
Eur. Phys. J. Plus \textbf{137} (2022) 590 
[arXiv:2112.11359 [nucl-th]].

\bibitem{Karapetyan:2020yhs}
G.~Karapetyan,
Eur. Phys. J. Plus \textbf{136} (2021) 122 
[arXiv:2003.08994 [hep-ph]].

\bibitem{Rougemont:2017tlu}
R.~Rougemont, R.~Critelli, J.~Noronha-Hostler, J.~Noronha and C.~Ratti,
Phys. Rev. D \textbf{96} (2017) 014032 
[arXiv:1704.05558 [hep-ph]].

\bibitem{Correa:2015vka}
R.~A.~C.~Correa and R.~da Rocha,
Eur. Phys. J. C \textbf{75} (2015)   522
[arXiv:1502.02283 [hep-th]].

\bibitem{Bazeia:2018uyg} D. Bazeia, D. C. Moreira and E. I. B. Rodrigues, J.\ Magn.\ Magn.\ Mater. {\bf 475} (2019)  734 [arXiv:1812.04950 [cond-mat.mes-hall]].


\bibitem{Stephens:2019tav}
M.~Stephens, S.~Vannah and M.~Gleiser,
Phys. Rev. D \textbf{102} (2020) 123514
[arXiv:1905.07472 [astro-ph.CO]].

\bibitem{Cruz:2018qby}
W.~T.~Cruz, D.~M.~Dantas, R.~V.~Maluf and C.~A.~S.~Almeida,
Annalen Phys. \textbf{531} (2019) 1900178
[arXiv:1810.03991 [gr-qc]].

\bibitem{Lee:2019tod}
C.~O.~Lee,
Phys. Lett. B \textbf{800} (2020) 135030
[arXiv:1908.06074 [hep-th]]. 

\bibitem{Barreto:2022ohl}
W.~Barreto and R.~da Rocha,
Phys. Rev. D \textbf{105} (2022)  064049
[arXiv:2201.08324 [hep-th]].


\bibitem{Gutsche:2017lyu}
T.~Gutsche, V.~E.~Lyubovitskij and I.~Schmidt,
Phys. Rev. D \textbf{97} (2018)  054011
[arXiv:1712.08410 [hep-ph]].

\bibitem{Gutsche:2019yoo}
T.~Gutsche, V.~E.~Lyubovitskij and I.~Schmidt,
Phys. Rev. D \textbf{101} (2020) 034026
[arXiv:1911.00076 [hep-ph]].


\bibitem{Drechsel:2007if}
D.~Drechsel, S.~S.~Kamalov and L.~Tiator,
Eur. Phys. J. A \textbf{34} (2007), 69-97
doi:10.1140/epja/i2007-10490-6
[arXiv:0710.0306 [nucl-th]].

\bibitem{Ramalho:2020nwk}
G.~Ramalho and M.~T.~Pe\~na,
Phys. Rev. D \textbf{101} (2020) 114008
[arXiv:2003.04850 [hep-ph]].



\bibitem{Peet:1998wn}
A.~W.~Peet and J.~Polchinski,
Phys. Rev. D \textbf{59} (1999) 065011
[arXiv:hep-th/9809022 [hep-th]].

\bibitem{Brodsky:2007hb}
S.~J.~Brodsky and G.~F.~de Teramond,
Phys. Rev. D \textbf{77} (2008) 056007
[arXiv:0707.3859 [hep-ph]].

\bibitem{Gutsche:2012wb}
T.~Gutsche, V.~E.~Lyubovitskij, I.~Schmidt and A.~Vega,
Phys. Rev. D \textbf{87} (2013) 016017
[arXiv:1212.6252 [hep-ph]].

\bibitem{Contino:2004vy}
R.~Contino and A.~Pomarol,
JHEP \textbf{11} (2004) 058
[arXiv:hep-th/0406257 [hep-th]].

\bibitem{Hong:2006ta}
D.~K.~Hong, T.~Inami and H.~U.~Yee,
Phys. Lett. B \textbf{646} (2007) 165
[arXiv:hep-ph/0609270 [hep-ph]].

\bibitem{Henningson:1998cd}
M.~Henningson and K.~Sfetsos,
Phys. Lett. B \textbf{431} (1998) 63
[arXiv:hep-th/9803251 [hep-th]].

\bibitem{BoschiFilho:2012xr}
H.~Boschi-Filho, N.~R.~F.~Braga, F.~Jugeau and M.~A.~C.~Torres,
Eur. Phys. J. C \textbf{73} (2013)  2540
[arXiv:1208.2291 [hep-th]].

\bibitem{Branz:2010ub}
T.~Branz, T.~Gutsche, V.~E.~Lyubovitskij, I.~Schmidt and A.~Vega,
Phys. Rev. D \textbf{82} (2010) 074022
[arXiv:1008.0268 [hep-ph]].

\bibitem{Brodsky:2006uqa}
S.~J.~Brodsky and G.~F.~de Teramond,
Phys. Rev. Lett. \textbf{96} (2006) 201601
[arXiv:hep-ph/0602252 [hep-ph]].







\bibitem{Grigoryan:2007my}
H.~R.~Grigoryan and A.~V.~Radyushkin,
Phys. Rev. D \textbf{76} (2007) 095007
[arXiv:0706.1543 [hep-ph]].


\bibitem{Ramalho:2017pyc}
G.~Ramalho and D.~Melnikov,
Phys. Rev. D \textbf{97} (2018) 034037
[arXiv:1703.03819 [hep-ph]].

\bibitem{Konen:1989jp}
W.~Konen and H.~J.~Weber,
Phys. Rev. D \textbf{41} (1990) 2201.


\bibitem{Jido:2007sm}
D.~Jido, M.~Doering and E.~Oset,
Phys. Rev. C \textbf{77} (2008) 065207
[arXiv:0712.0038 [nucl-th]].

\bibitem{Tiator:2011pw}
L.~Tiator, D.~Drechsel, S.~S.~Kamalov and M.~Vanderhaeghen,
Eur. Phys. J. ST \textbf{198} (2011) 141
[arXiv:1109.6745 [nucl-th]].


\bibitem{Aznauryan:2011qj}
I.~G.~Aznauryan and V.~D.~Burkert,
Prog. Part. Nucl. Phys. \textbf{67} (2012) 1
[arXiv:1109.1720 [hep-ph]].




\bibitem{Aznauryan:2012ec}
I.~G.~Aznauryan and V.~D.~Burkert,
Phys. Rev. C \textbf{85} (2012) 055202
[arXiv:1201.5759 [hep-ph]].

\end{thebibliography}
\end{document}